\documentclass[english,twocolumn,superscriptaddress]{revtex4-1}

\usepackage{amsmath}

\usepackage[T1]{fontenc}
\usepackage[utf8]{inputenc}
\usepackage{graphicx}
\usepackage{xcolor}
\usepackage[caption=false]{subfig}
\usepackage{newtxtext,newtxmath}

\usepackage{bbold}
\newcommand{\expv}[1]{\left\langle #1\right\rangle}
\newcommand{\eq}[1]{Equation~(#1)}
\newcommand{\fig}[1]{Figure~#1}

\newcommand{\identity}{\mathbb{1}}

\begin{document}

\title{Modular hierarchical and power--law small--world networks bear
structural optima for minimal first passage times and cover time}

\author{Benjamin F. Maier}
\email{Corresponding author: bfmaier@physik.hu-berlin.de}
\affiliation{Robert Koch Institute, Nordufer 20, D-13353 Berlin, Germany}
\affiliation{Department of Physics, Humboldt Universit\"at zu Berlin, Newtonstr. 15, 
        D-12489 Berlin, Germany}
\author{Cristi\'an Huepe}
\affiliation{Huepe Labs, 2713 West Haddon Ave, 
        Chicago, Illinois 60622, USA}
\affiliation{Northwestern Institute on Complex Systems \& ESAM,
        Northwestern University, Evanston, Illinois 60208, USA}

\author{Dirk Brockmann}
\affiliation{Robert Koch Institute, Nordufer 20, D-13353 Berlin,
    Germany}
\affiliation{Institute for Theoretical Biology, Humboldt Universit\"at zu
        Berlin, Philippstr. 13, D-10115 Berlin, Germany}

\date{\today}

\begin{abstract}
    \begin{center}
{\bfseries Abstract}
\end{center}
Networks that are organized as a hierarchy of modules have been the
subject of much research, mainly focusing on algorithms that can extract
this community structure from data. The question of why modular hierarchical
organizations are so ubiquitous in nature, however, has received less
attention. One hypothesis is that modular hierarchical topologies
may provide an optimal structure for certain dynamical processes. We
revisit a modular hierarchical network model that interpolates, using
a single parameter, between two known network topologies: from strong hierarchical 
modularity to an Erdős--Rényi random connectivity structure. We show that
this model displays a similar small--world effect as the Kleinberg
model, where the connection probability between nodes decays algebraically
with distance. We find that there is an optimal structure, in both
models, for which the pair--averaged first passage time (FPT) and
mean cover time of a discrete--time random walk are minimal, and
provide a heuristic explanation for this effect. Finally, we show
that analytic predictions for the pair--averaged FPT based on an effective medium
approximation fail to reproduce these minima, which implies that
their presence is due to a network structure effect.
\\
Keywords: modular hierarchical networks, Kleinberg networks, small--world networks, random walks, first passage time, cover time, effective medium approximation
\end{abstract}

\keywords{modular hierarchical networks, Kleinberg networks,
    small--world networks, random walks, first passage time, cover time, effective medium approximation}

\maketitle
\section{Introduction}

In the last two decades, there has been much progress in the study
of networks that display community structures, also referred to as
modules \cite{fortunato_community_2010,lancichinetti_detecting_2009,albert_statistical_2002,newman_modularity_2006}.
Various works have also considered the presence of hierarchies
of these communities, describing networks that contain further communities
within communities \cite{fortunato_community_2010,sales-pardo_extracting_2007,clauset_hierarchical_2008,peixoto_hierarchical_2014,rosvall_multilevel_2011}.
It is also widely known that a remarkable number of natural and social
systems can be described as a hierarchy of modules \cite{simon_architecture_1962,peixoto_hierarchical_2014},
such as brain networks \cite{meunier_hierarchical_2009,meunier_modular_2010,kaiser_limited_2010,robinson_dynamical_2009,sarkar_spectral_2013,klimm_resolving_2014},
metabolic and cell networks \cite{clauset_hierarchical_2008,ravasz_hierarchical_2002,barabasi_network_2004},
human transport networks \cite{sales-pardo_extracting_2007,yerra_emergence_2005},
ecological systems \cite{clauset_hierarchical_2008,smith_hierarchical_2014,webster_hierarchical_1979}
and social networks \cite{sarkar_spectral_2013,simon_architecture_1962}.
However, although a significant amount of research has focused on
inferring the hierarchical structure contained in these real-world
networks \cite{sales-pardo_extracting_2007,clauset_hierarchical_2008,peixoto_hierarchical_2014,rosvall_multilevel_2011,sarkar_spectral_2013,arenas_synchronization_2006},
fewer efforts have addressed the question of why such hierarchical organization
has emerged and persisted in the first place \cite{smith_hierarchical_2014,yerra_emergence_2005,pan_modular_2008}.
One potential reason could be that hierarchical networks provide an
optimal interaction topology, e.g.~increase a system's structural or dynamic stability and robustness \cite{smith_hierarchical_2014,rao_dynamic_2013,kaiser_limited_2010,pan_modular_2008}.
It was shown, in particular, that certain classes of modular hierarchical
(MH) networks described in a model by Watts \textit{et al.}~\cite{watts_collective_1998,watts_identity_2002}
provide an optimal structure for reaching any specific node from any
other in a minimal number of steps, using only local information.
This has served as an insight for understanding the outcome of Milgram's
famous ``small--world'' experiment \cite{travers_small_1967}. While
suggestive of the unique properties of MH networks, this locally-informed
search process does not necessarily reflect the underlying dynamics
of many natural hierarchical systems, which could follow more random,
diffusion--like processes.

In this paper we study the topological structure of and random walk processes
on a class of MH networks with different levels of hierarchical
modularity. This class corresponds to a variant of Watts' hierarchical
model, in which all hierarchical levels are statistically
self--similar and where the amount of hierarchical modularity can be
varied using a single structural control parameter, while keeping
the mean degree fixed. We will refer to this model as the \textit{self--similar 
modular hierarchical} (SSMH) network model. In this model, the control
parameter determines the degree of topological non--locality of the
system, i.e.~the fraction of connections that are made within
a module at each level of the hierarchy. We also analyze the class
of networks resulting from a variant of a one--dimensional Kleinberg
small--world model \cite{kleinberg_small--world_2000},
in which all pairs of nodes including nearest neighbors are linked with
power--law probability on the distance between them. The mean degree
is also kept constant and the structural control parameter is defined
as the exponent of the power--law, thus controlling the level of spatial
non--locality in a one--dimensional embedding space. We will refer
to this model as the \textit{power--law small--world} (PLSW) network model. We will show
that the SSMH and PLSW models share many properties, although the
latter does not have an inherent modular structure.

The definition of the SSMH and PLSW models as statistically self--similar
systems, each with a single structural control parameter that determines
the level of non--locality of the connections, allows us to define
different topological phases, ranging from a network with only local
connections to an Erdős--Rényi network with fully random connections
that span the whole system, and to find analytical expressions for
the critical parameter values that separate them. It will also allow
us to find an explicit relationship between both control parameters
and to derive several other analytical expressions, such as the SSMH
model's degree variance, generating function, and clustering, quantities that we explicitly derive in Appendix \ref{app:SSMH_properties}.

We analyze the dynamics of discrete--time random walks
on SSMH and PLSW systems, for different values of their control
parameters. We first compute the mean first passage time (FPT) of a random
walk between two nodes, averaged over all possible pairs of nodes.
This gives a measure of the characteristic diffusion times on both
types of networks and corresponds to an upper bound for the averaged
locally--informed search times of general search processes in these
networks, which was computed in \cite{newman_networks:_2010}. We also find
the mean cover time, the mean number of steps it takes
for a random walker to visit all the nodes in the network. This gives,
in turn, an upper bound for the time required for exhaustive search
processes in these networks. The main observation that results from
these computations is that both quantities display a minimum at intermediate
levels of their structural control parameters. We note that
these minimum times only emerge when computed on actual network
realizations, and not when using an effective medium approximation.
In this approximation, a link between two nodes is replaced by a transition probability in an effectively continuous medium that captures the topological distances between nodes in the network
\cite{bruggeman_berechnung_1935,thiel_effective-medium_2016}.
The emergence of these minima must thus be 
the result of changes in the network structure and cannot
be explained using averaged dynamics. Finally, we provide an argument
to understand the presence of these minima analytically, as the result
of two opposing effects: a decrease in the local clustering and increase in
the inverse degree variance. We use this argument to derive a heuristic
expression for the lower bound of the pair--averaged FPT in networks
with non--zero local clustering.

The paper is organized as follows. In Section~\ref{sec:models}, we introduce the SSMH and PLSW network models and analyze their structural properties for different levels of hierarchical clustering.
In Section~\ref{sec:random_walk}, we consider discrete--time random walk dynamics, computing the pair--averaged FPT and mean cover time of random walkers for different levels of hierarchical clustering. 
By performing numerical network realizations, we find that these two characteristic times display a minimum as a function of the structural control parameter for both network models. 
We then show that these minima are lost if we use an effective medium approximation to compute these characteristic times analytically, since the pair--averaged FPT and the mean cover time monotonically decrease as the network is made more homogeneously random by increasing the structural control parameter. The minima are instead heuristically explained by a concurrent decrease of local clustering and increase of node degree heterogeneity.
Finally, Section~\ref{sec:discussion} presents our discussion and conclusions.
In addition, the Appendices provide detailed calculations on the properties of SSMH and PLSW networks, as well as on our effective medium and heuristic analyses. Appendix E includes a comparison to the original Watts--Strogatz small--world network model, which also displays a minimum in the pair--averaged FPT.

\section{Network models}

\label{sec:models}

We begin by describing the two classes of models that we will study in this paper. Although both are similar to network generating algorithms that have previously been introduced in the literature, their original purpose was different from the study of diffusive dynamics on MH structures that is our focus here. We thus introduce in this Section variants of these models that are fully self--similar, which result in simpler analytical expressions that help us better understand the properties of the resulting networks.

\subsection{Self--similar modular hierarchical (SSMH) networks}

\label{sec:mhrn_model}

We present first a stochastic block model that can generate self--similar random networks with different degrees of hierarchical modularity; the SSMH model. This is a variant of a model introduced by Watts \textit{et al.}~\cite{watts_identity_2002} to describe search approaches in social networks.

\begin{figure*}
\subfloat[\label{fig:construction_group_layers}]{\includegraphics[width=6.5cm]{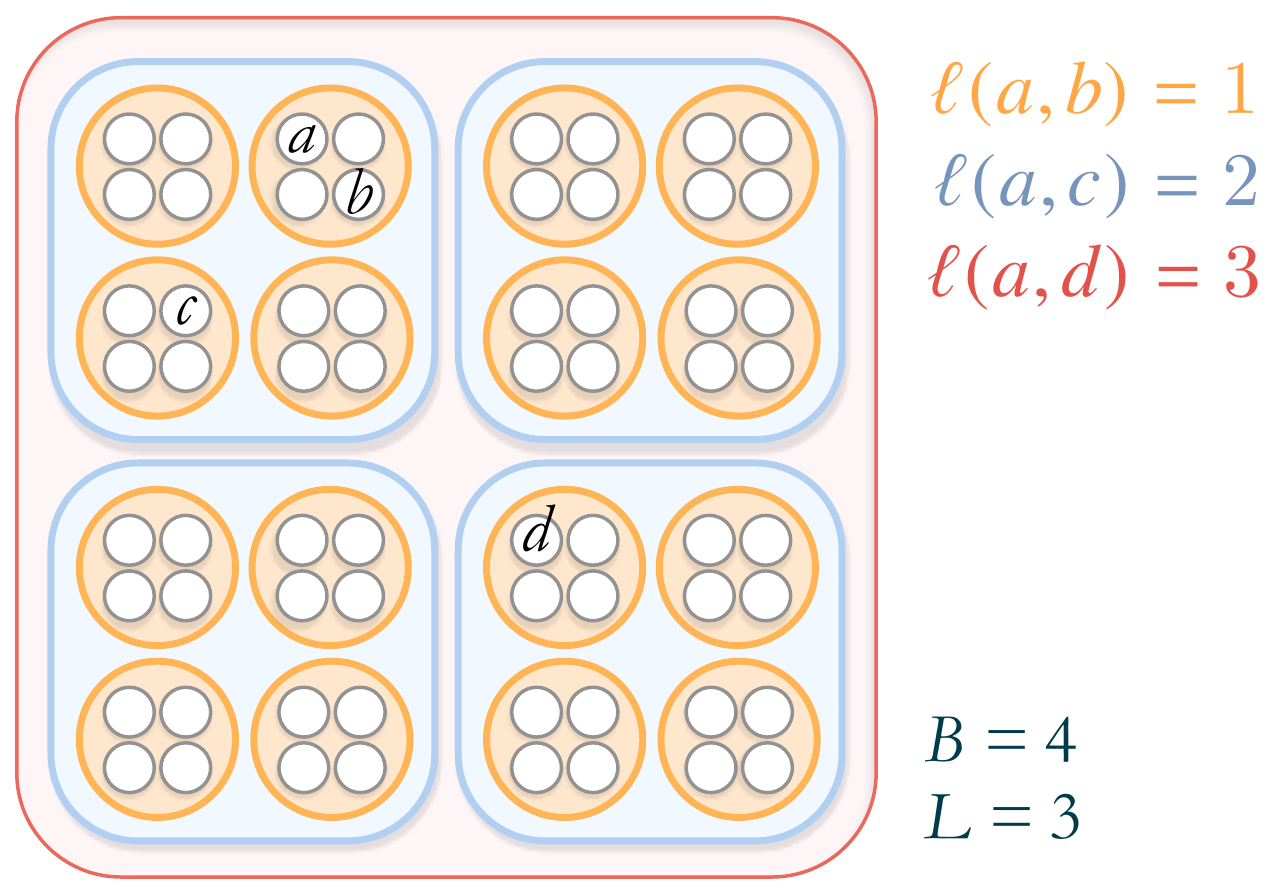}

}\hfill\subfloat[\label{fig:construction_tree}]{\includegraphics[width=4.5cm]{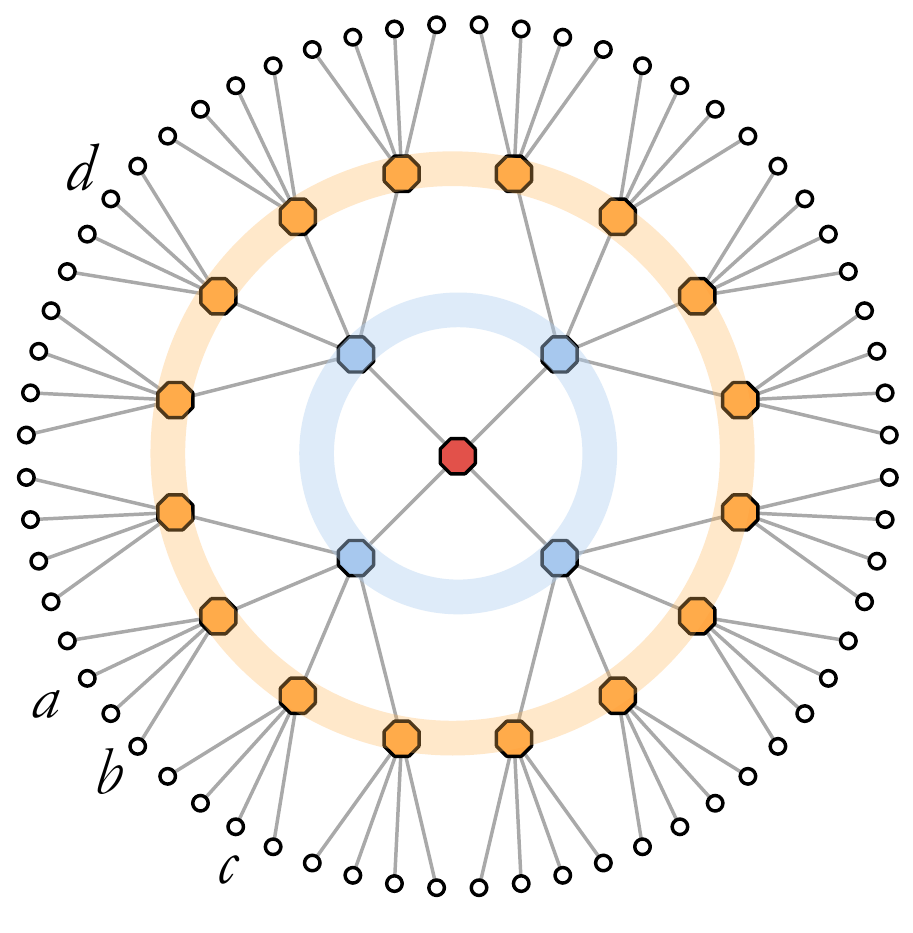}

}\hfill\subfloat[\label{fig:construction_adjacency}]{\includegraphics[width=4.5cm]{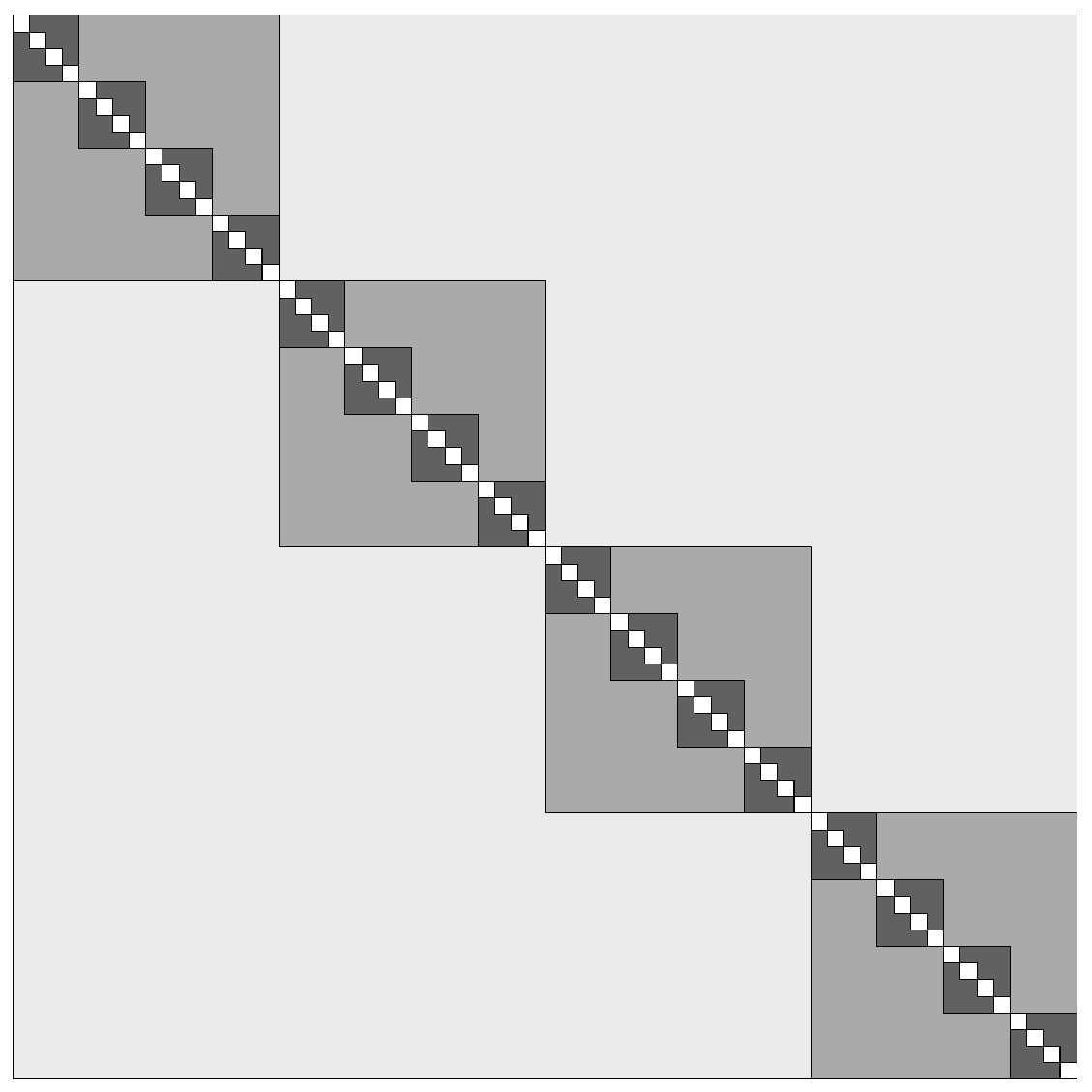}

} \caption[width=\textwidth]{\label{fig:construction} Different diagrams
    illustrating the hierarchical modular structures studied in this
    paper, using a case with number of modules per level (base) $B=4$
    and total number of levels $L=3$ as an example. Panels \textbf{(a)}
    and \textbf{(b)} represent the hierarchy of modules as embedded
    structures and a hierarchical tree, respectively. The network nodes
    (white circles) are grouped into modules of $B$ components at level
    1 (orange), which are in turn grouped into $B$-sized modules
    (containing $B^{2}$ nodes) at level 2 (blue). Finally, these are
    grouped into a single level 3 structure (red) that also contains $B$
    submodules, and thus $B^{3}$ nodes in total. The nodes labeled $a$
    and $b$   belong to the same level-1 module, those labeled $a$ and
    $c$ are in the same level-2 module (but not the same level-1
    moduls), and those labeled $a$ and $d$, in the same level-3 module
    (but not the same lower-level ones). A hierarchical modular network
    can then be built by linking with lower probability the nodes that
    belong to higher level modules only. Panel \textbf{(c)} displays the
    structure of the resulting adjacency matrix, with darker shades of
    grey representing a higher probability of having '1' entries that denote connections.}
\end{figure*}

\subsubsection{Structure}

The model assumes that each node belongs to a peer group (often referred to as a module or community) that is part of a bigger community of modules, which is in turn part of a yet larger community of modules of modules, etc., as depicted in Figure~\ref{fig:construction_group_layers}.
This structure of communities and subcommunities can be represented by an underlying hierarchical tree (see Figure~\ref{fig:construction_tree}), where we define the \textit{hierarchical distance} $\ell$ between two nodes as the smallest number of levels that we need to traverse up the tree to find a common ancestor. Note that, in this picture, only the leaves represent actual nodes in the generated network while the rest of the tree is only used to define a MH connectivity structure. We thus define a network where the probability of having a connection between two nodes decreases as their distance $\ell$ increases. This allows us to generate a network structure composed of a hierarchy of modules, where each can be more connected internally than externally. 

For simplicity, we consider here self--similar MH structures, where all modules have the same number of submodules. We thus generate our SSMH networks starting from a $B$-ary tree of height $L$, where $B$ is the module size and $L$ is the total number of hierarchical layers.
Then the final network consists of $N=B^{L}$ nodes, which could potentially be connected to up to 
\begin{equation}
k_{\ell}^{\max}=B^{\ell}-B^{\ell-1}=B^{\ell-1}(B-1)\label{eq:k_l_max}
\end{equation}
other nodes of hierarchical distance $\ell$. 
Although this number of potential links to other nodes grows exponentially with $\ell$, in order to generate MH structures we would like the actual number of connections to nodes in other modules to be typically smaller than but of similar order as the number of connections to nodes within the same module. We thus define a connection probability $p_{\ell}$ that decreases exponentially with increasing hierarchical distance, given by
\[
p_{\ell}\propto\left(\frac{\xi}{B}\right)^{\ell-1}.
\]
Here, $0\leq\xi\leq B$ is defined as the MH structural control parameter. By choosing different values of $\xi$, we can thus generate a class of networks with different degrees of hierarchical modularity. Since the mean degree of a network strongly affects its properties, we will keep it constant for all members of the class, in order to properly compare them. To do this, we first find the mean degree of each node with respect to all other nodes at hierarchical distance $\ell$, which is given by $\expv{k_{\ell}}=p_{\ell}k_{\ell}^{\max}$. We then compute the mean degree of all nodes $\expv k=\sum_{\ell=1}^{L}\expv{k_{\ell}}$ and use it to normalize the total number of connections, obtaining
\begin{equation}
p_{\ell}=\frac{\left\langle k\right\rangle }{B-1}\left(\frac{1-\xi}{1-\xi^{L}}\right)\left(\frac{\xi}{B}\right)^{\ell-1}.\label{eq:p_l}
\end{equation}
Note that if $\expv{k} > B-1$, the MH structural control parameter $\xi$ can only be chosen to be larger than or equal to $\xi_{\min} > 0$, where $(B-1)\left(1-\xi_{\min}^{L}\right)=\left\langle k\right\rangle (1-\xi_{\min})$, in order to have a connection probability that satisfies  $p_{\ell}\leq1$ for all hierarchical distances $\ell \geq 1$.
\subsubsection{Hierarchy phases}
\label{sec:network_phases}
We now describe the different topological phases of the SSMH model as a function of the hierarchical modularity structural control parameter $\xi$, while keeping $B$, $L$ and $\expv k$ fixed.

\begin{figure*}
     \centering
     \includegraphics[width=\textwidth]{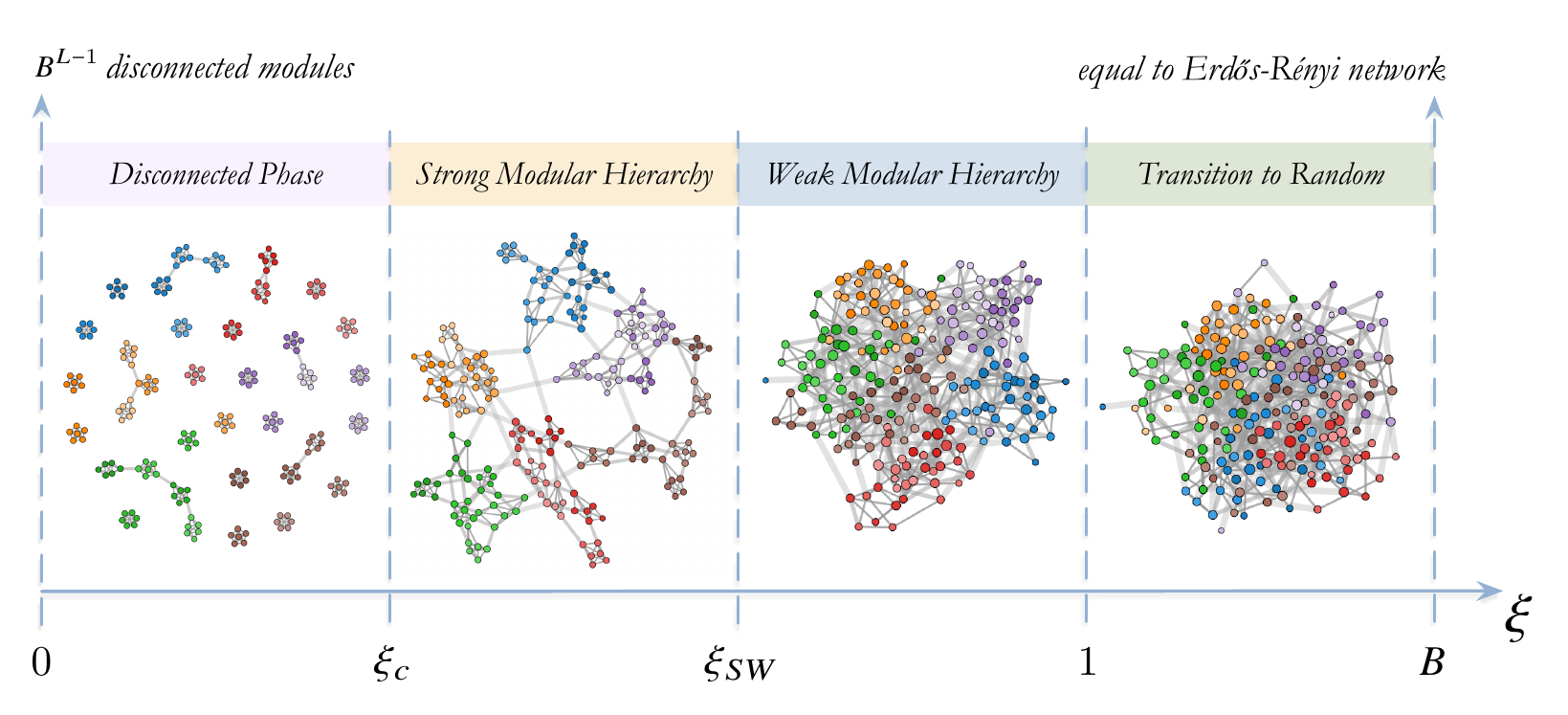} 

     \caption{\label{fig:phases} 
    Topological phases of the SSMH model as a function of  the structural control  parameter $\xi$ that determines the degree of hierarchical modularity of the network (see Section~\ref{sec:network_phases}).}
\end{figure*}
If we select parameters for which $\xi_{\min}=0$, in the $\xi=0$ case the network consists of $B^{L-1}$ densely connected graphs of $B$ nodes.
As $\xi$ is increased, links are redistributed from the lowest hierarchical layer to higher ones, while keeping the total number of links constant. The system thus goes through the following phases, depicted in \fig{\ref{fig:phases}}.
\begin{itemize}
\item $\xi\gtrsim\xi_{c}\,:$ A large component of size $\mathcal{O}(B^{L})$
emerges.
\item $\xi\leq\xi_{SW}\,:$ The network is in a phase with \textit{strong} hierarchical modularity, where an average node has more links to nodes in its own lowest hierarchy module (level $\ell=1$) than to nodes in all other hierarchy layer groups $\ell>1$ combined, satisfying
\begin{equation}
\expv{k_{1}}\geq\expv{k_{2}}+\cdots+\expv{k_{L}}.\label{eq:strong_hierarchy}
\end{equation}
The value of $\xi_{SW}$ is given by the solution to equation 
$0=\xi_{SW}^{L}-2\xi_{SW}+1$ (with $1/2\leq\xi_{SW}\leq1$), which quickly approaches $\xi_{SW}=1/2$ as $L$ is increased. 
\item $\xi_{SW}<\xi\leq1\,:$ The network is in a phase with \textit{weak} hierarchical modularity, where the degree at each hierarchical level is smaller than the degree at the next one, that is
\begin{equation}
\left\langle k_{1}\right\rangle \geq\left\langle k_{2}\right\rangle \geq\dots\geq\left\langle k_{L}\right\rangle \label{eq:degree_hierarchy}
\end{equation}
is satisfied, but the condition in \eq{\ref{eq:strong_hierarchy}} is not.\item $1<\xi<B\,:$ In this phase, only the basic hierarchical modularity probability condition
\begin{equation}
p_{1}>p_{2}>\ldots>p_{L}.\label{eq:prob_hierarchy-1}
\end{equation}
is satisfied, as the network transitions to a homogeneous random structure.
\item $\xi=B\,:$ The network is identical to an Erdős--Rényi random network,
since $p_{\ell}$ is constant for all layers $\ell$. 
\end{itemize}

\subsubsection{Self--similar modular hierarchical network generating algorithm}
\label{sec:mhrn_generation_algorithm}

In order to efficiently generate SSMH topologies, we implemented an $\mathcal{O}(\expv m)$ algorithm (where $\expv{m}$
is the mean total number of edges), which we briefly explain as follows. We begin
by constructing $B^{L-1}$ Erdős--Rényi networks of size $B$ and with connection
probability $p_{1}$, using the algorithm described in \cite{batagelj_efficient_2005}.
Those will be the base modules in layer $1$. Subsequently, for every
layer $\ell>1$, we draw a number $m_{\ell}$ of edges in this layer
from a binomial distribution with parameters $m_{\ell}^{\max}=B^{L}B^{\ell-1}(B-1)/2$
and $p=p_{\ell}$. For every edge appearing in the layer, we pick
a random node $u$ from the set of all nodes, and a second node $v$
from all $B-1$ modules that this node can reach in this layer. If there
is not yet an edge connecting $u$ and $v$, the edge is assigned, 
otherwise a new originating node $u$ is picked. 

We developed a custom Python/C++/Matlab package that produces SSMH topologies in this manner publicly available for download at \cite{maier_cmhrn_2018}.

\subsection{Power--law small--world (PLSW) networks}

\label{sec:kleinberg_model}

Since modular hierarchical networks have been shown to display small--world properties \cite{watts_collective_1998}, it will be interesting to compare their diffusive dynamics to those on a small--world system. To this end, we will consider a variation of a model established by Kleinberg~\cite{kleinberg_small--world_2000}, which we will use to generate PLSW network topologies. 
In the original Kleinberg model, $N$ nodes are embedded in a low--dimensional space and connected to their nearest neighbors. Additional long--range links are then added, with a probability $P$ that decays as a power--law with the lattice distance $n$ between the two nodes, following $P\propto n^{-\kappa}$, with $\kappa>0.$
In this paper, we consider a variant of this model where all nodes are linked with the same probability distribution, which decays as a power--law with their one-dimensional distance in an embedding space, and neighbors are not automatically connected. This PLSW model can be described as follows. Given two distinct nodes $u$ and $v$ with indices $0\leq i_{u}\leq N-1$ and $0\leq i_{v}\leq N-1$, we define the smallest distance between them in periodic boundary conditions as
\[
n(u,v)=\min(|i_{v}-i_{u}|,\ N-|i_{v}-i_{u}|).
\]
In our PLSW model, 
the connection probability between any two nodes is then given by 
\[
c_{n(u,v)}=p_{0}|n|^{\mu-1}.
\]
Here, $-\infty < \mu \leq 1$ is our structural control parameter that determines, in this case, the degree of non--locality in the connections (larger $\mu$ values imply higher long--range connection probability) and $p_{0}$ is a normalization constant. 
With these definitions, we can then establish a direct relationship between the probability of linking two nodes at a given distance in our SSMH and PLSW models. As derived in Appendix \ref{sec:mapping_mhrn_to_kleinberg}, this leads us to the expression  
\[
\mu=\log\xi/\log B,
\]
which relates the structural control parameters, $\xi$ and $\mu$, of these two models. Again, a constant node degree is important for comparing the dynamics on this class of networks. We therefore normalize $c_{n}$ by selecting $p_0$
such that it sums to the node degree imposed in Equation (\ref{eq:kleinberg_norm}) of Appendix \ref{sec:mapping_mhrn_to_kleinberg}.

\begin{figure}
\centering
\includegraphics[width=3.375in]{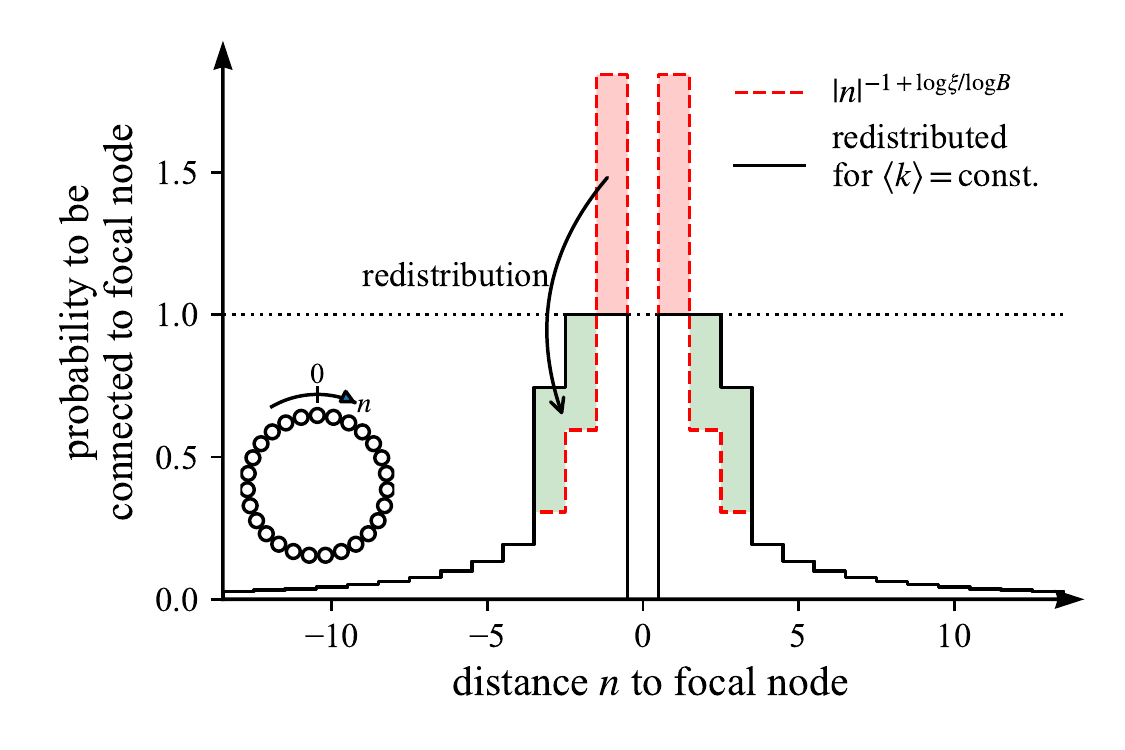}\,\caption{\label{fig:mod_Kleinberg_pmf}
Method for redistributing the probability mass function of the PLSW network model defined in this paper (as a variation of the Kleinberg small--world model) so that it does not exceed unity in numerical calculations. The original connection probability (red dashed line) is modified by redistributing the excess probability to its nearest neighbors. }
\end{figure}

We note that our definition of the PLSW model above will produce a problem for large negative values of $\mu$, for which the connection probability between close neighbors will exceed unity. To avoid having probability values that are larger than one, we redistribute the excess probability to the nearest neighbors until we run out of excess probability, as illustrated in \fig{\ref{fig:mod_Kleinberg_pmf}}. This has the net effect of producing a $\expv k$-nearest neighbor lattice for $\mu\ll0$ ($\xi\rightarrow0$, respectively), similar to  that in the original Watts' and Strogatz's small--world model \cite{watts_collective_1998}.
We find the following approximate expression, that can be used to compute the critical value $\mu_{c}$ at which the probability to connect to the nearest neighbor exceeds unity
\begin{equation}
1\approx\frac{\expv k}{2}\left(\int\limits _{1}^{N/2}\mathrm{d}n\frac{1}{n^{\mu_{c}-1}}\right)^{-1}=\expv k\frac{2\mu_{c}-4}{4-2^{\mu_{c}}N^{2-\mu_{c}}}.\label{eq:critical_mu}
\end{equation}

\begin{figure*}
\centering
\includegraphics[width=\textwidth]{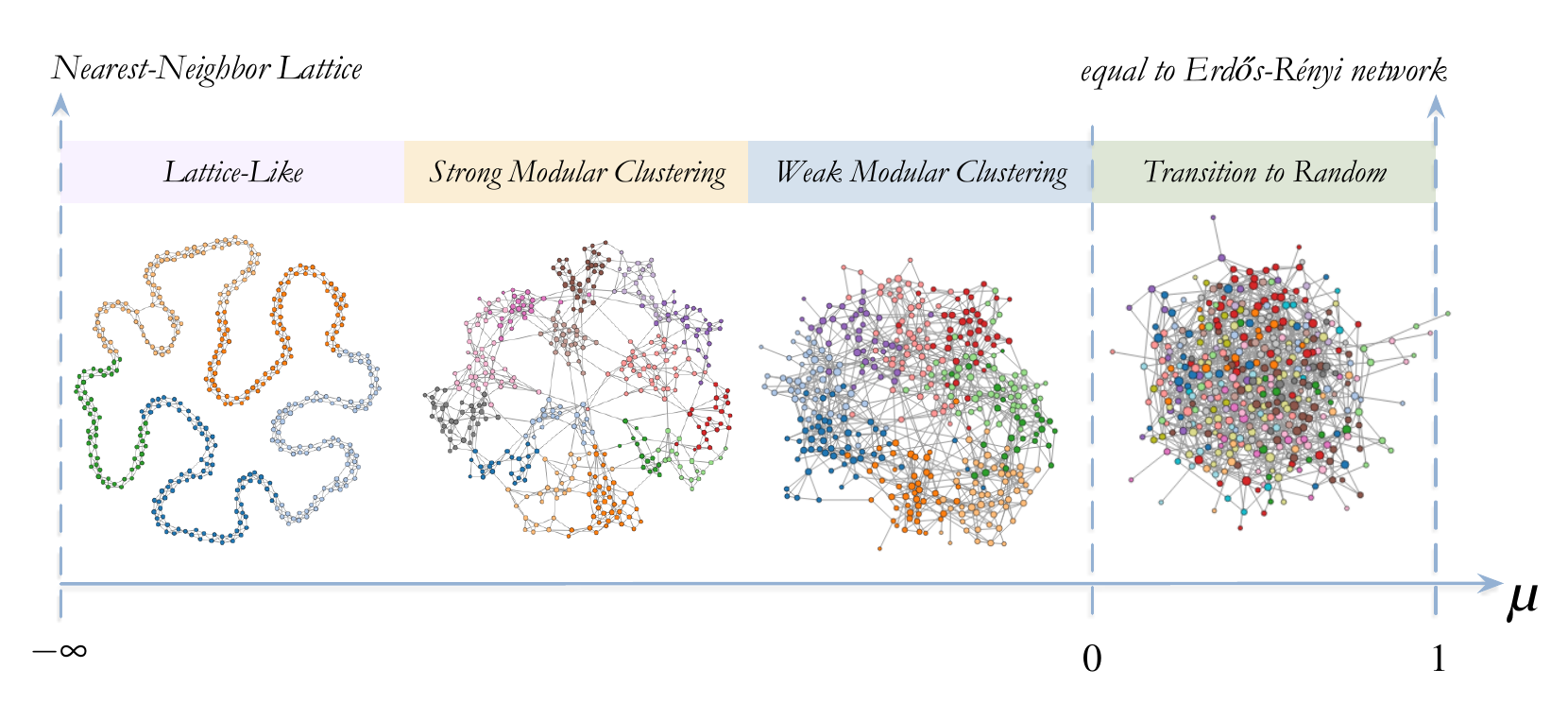}\caption{\label{fig:Kleinberg_phases}
Topological phases of the PLSW model as a function of the structural control parameter $\mu=\log\xi/\log B$ that determines the degree of non--locality of the connections (see Section~\ref{sec:kleinberg_model}).}
\end{figure*}

Figure~\ref{fig:Kleinberg_phases} illustrates the different topological phases that can be observed in the class of networks generated by the PLSW model, in relation to the corresponding phases of the SSMH model. In both cases, as the structural control parameter ($\xi$ and $\mu$, respectively) is decreased, the topology changes from that of a homogeneous Erdős--Rényi random graph to a network with stronger local interactions. However, the type of local structures that emerge in both cases is very different.  Whereas in the SSMH model the increase in local interactions leads to a stronger hierarchy of modules, in the PLSW case it results in chains of nearest neighbors. Therefore, for very low values of the structural order parameters ($\xi$ and $\mu$) the SSMH model produces dense, disconnected clusters while the PLSW model produces chains that resemble nearest neighbor lattices.



\section{Discrete--time random walks}
\label{sec:random_walk}
In this section, we investigate discrete--time random walks on the SSMH and the PLSW models. We will compute the global mean first passage time (FPT) and the mean cover time on different topologies. This will provide a measure of the efficiency of  diffusive dynamics as a model for random search processes on these networks.
\subsection{Theory}
\label{sec:random_walk_theory}

We begin by considering an undirected network with adjacency
matrix $A_{vu}$, where $A_{vu}>0$ if nodes $u$ and $v$ are connected
and $A_{vu}=0$ if not. On this network, a random walk is a Markov
process where a walker which resides on node $u$ at time $t$ has transition
probability $W_{vu}=A_{vu}/k_{u}$ to be on node $v$ at time $t+1$.
It is thus governed by the master equation 
\[
P_{v}(t+1)=\sum_{u=1}^{N}W_{vu}P_{u}(t).
\]
This process will approach the equilibrium  distribution $P_{v}^{\star}=k_{v}/2m$
with $m$ being the number of edges. 

The first quantity that we will study is the mean FPT between
two nodes $u$ and $v$, defined as the mean number of steps $\tau_{vu}$ it takes
for a random walker starting at a node $u$ to visit a target node $v$ for the first time.
In order to consider a first passage measure that is independent from the starting node, we define the global mean FPT of a target node $v$ as the average first passage time when starting from any other node in the network, that is
\[
\tau_{v}=\frac{1}{N-1}\sum_{u=1,u\neq v}^{N}\tau_{vu}.
\]
Nodes with a small global mean FPT are thus typically visited earlier and more often
than nodes with a high global mean FPT. Furthermore, in order to define a first passage measure that estimates the typical relaxation timescale of a random diffusion process throughout the whole network (independently of the specific target node considered), we compute the global mean FPT averaged over all target nodes in the network, given by
\[
\expv\tau=\frac{1}{N}\sum_{v=1}^{N}\tau_{v}.
\]
This quantity is identical to the FPT averaged over all pairs, given by
\[
\expv\tau=\frac{1}{N(N-1)}\sum_{u=1}^{N}\sum_{v\neq u}^{N}\tau_{vu}.
\]
In addition to this pair--averaged FPT, we will also investigate the mean cover time $\expv T$, given by the mean number of steps it takes for a single walker to visit all other nodes at least once, averaged over all possible starting positions. 
This quantity can be viewed as the typical time it takes for a randomly diffusing process or signal to reach the whole network. This is an important timescale since it captures the time it takes to locate a target node in a network reliably.
%
\subsection{Numerical results}
\label{sec:numerical_results}

\begin{figure*}
\begin{centering}
\subfloat{\includegraphics[width=9cm]{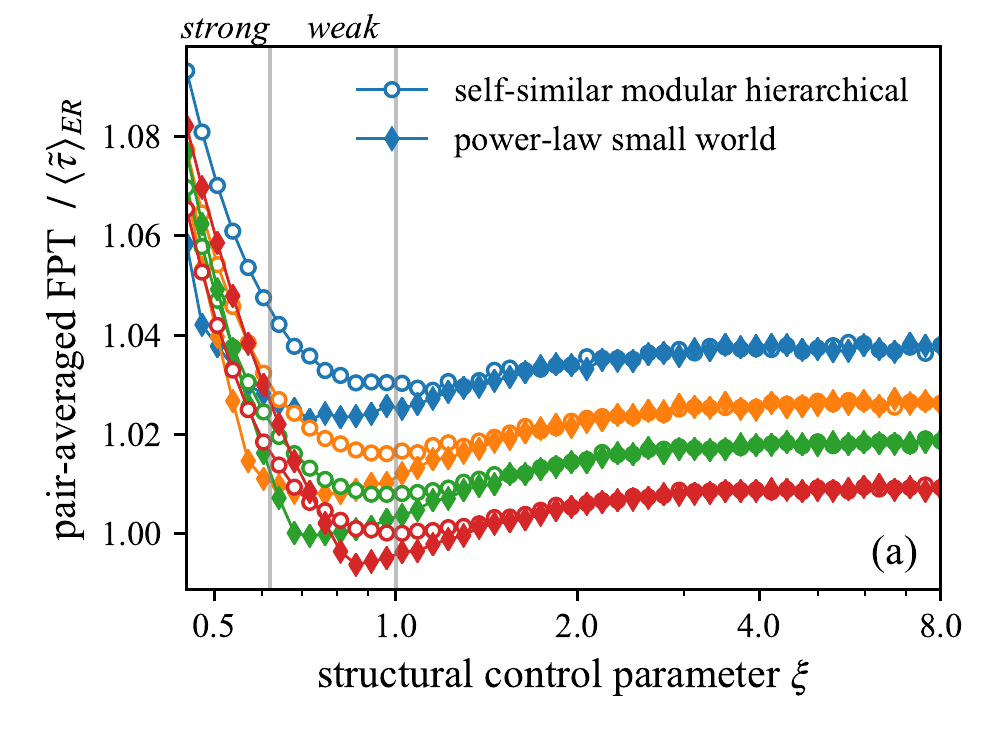}}\subfloat{\includegraphics[width=9cm]{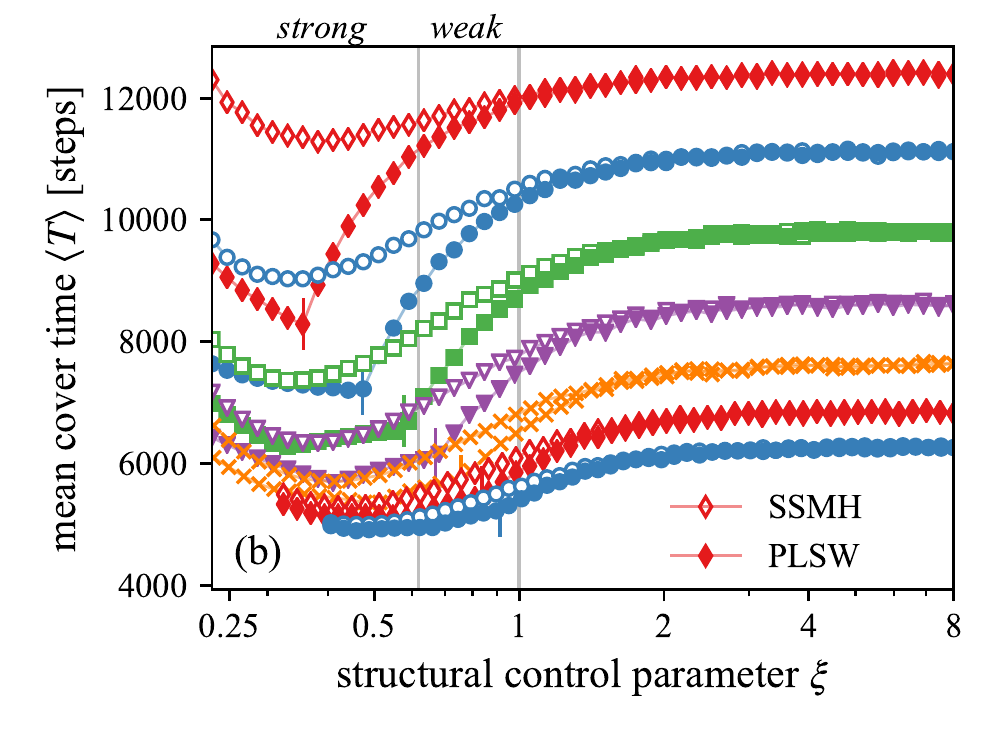}}
\par\end{centering}
\raggedright{}\caption{
    \label{fig:SSMH_MFPT}
    Characteristic diffusion time measures computed numerically for the self--similar modular hierarchical (SSMH) and the power--law small world (PLSW) network models considered in this paper.
    \textbf{(a)} Pair--averaged FPT as a function of the structural control parameter $\xi$ for $B=8$, $L=3$, and $\expv k\in\{6,7,8,10\}$ (top to bottom curves, respectively),  averaged over 2000 SSMH and PLSW network realizations and normalized by the lower bound of the pair--averaged FPT on Erdős--Rényi networks given in Equation (\ref{eq:heuristic_clustered_MGMFPT}).
    \textbf{(b)} Corresponding mean cover time for networks with $\expv k\in\{5,6,7,8,9,10,11\}$ (top to bottom curves, respectively). 
    For PLSW cases with small mean degree $\expv k$, the $\xi$ value of the minimum mean cover time coincides with the critical $\mu_{c}$ (marked by a short vertical line) obtained from Equation (\ref{eq:critical_mu}). 
    This $\mu_{c}$ corresponds to the structural control parameter value below which the connection probability to a focal node's nearest neighbors exceeds one, and must therefore be redistributed to other nearby neighbors.
    In this regime, the existence of the observed sharp minimum could thus be related to the enhanced non--local connectivity that must be imposed for $\mu < \mu_{c}$. 
    However, $\mu_{c}$ does not coincide with the observed minimum for larger $\expv k$ values, so its presence cannot be directly related to the link probability redistribution process in all regimes.
}
\end{figure*}

We begin by computing the pair--averaged FPT and the mean cover time numerically on actual network realizations. To this end, we set $B=8$, $L=3$, and $\expv k\in\{5,6,7,8,9,10,11\}$ and scan different values of the structural control parameter $0.25<\xi\leq B$ (or the corresponding $\mu$ parameter).

For each parameter set, we generated $2000$ different SSMH network realizations with $N=B^{L}$ nodes, using the algorithm described in Section \ref{sec:mhrn_generation_algorithm}. 
Similarly, we generated $2000$ PLSW network realizations with $N=B^{L}$ nodes, the same mean degree $\expv k$ values, and the corresponding structural control parameter $\mu=\log\xi/\log B$, using the algorithm described in Section \ref{sec:kleinberg_model}.
After selecting the largest component of each of the resulting networks, we measured the pair--averaged FPT $\expv\tau$ using the eigenvalues and eigenvectors of the unnormalized graph Laplacian, as described in \cite{lin_mean_2012}. To compute the mean cover time $\expv T$, we simulated discrete--time random walks on each network, starting with a single random walker placed on each node of the network. For each walker, we obtained a cover time as the number of steps it took to visit all nodes at least once. Then, $\expv T$ was computed as the mean over all walkers and over all network realizations. 
Simulations were performed using a custom Python/C++ package publicly available for download \cite{maier_cnetworkdiff_2017}. 

We observe in \fig{\ref{fig:SSMH_MFPT}a} that 
a minimum in the pair--averaged FPT emerges in the $\xi\lesssim 1$ region for, both, 
the SSMH and the PLSW networks. This corresponds to the weak hierarchical
clustering regime, where an average node has more connections to nodes in lower hierarchical layers than to nodes in higher layers, but where both quantities are of similar order.
%
\fig{\ref{fig:SSMH_MFPT}b} shows that the mean cover time for SSMH and PLSW numerical network realizations also displays a minimum. Here the minimum is in the strong hierarchical clustering phase for small $\expv k$ values and shifts towards the weak hierarchical clustering regime as $\expv k$ is increased.
We note that, for PLSW networks with low mean degree $\expv k$ values, this minimum coincides with the critical structure parameter $\mu_{c}$ below which the connection probability to a focal node’s nearest neighbors would exceed one, and was therefore redistributed to other nearby neighbors.
This correspondence is lost for larger $\expv k$ values, however, showing that this minimum does not have to be produced by this potentially spurious effect. 
We also note that the fact that we observe a minimum in the strong hierarchical clustering regime is surprising, since networks are considerably more clustered in this regime than, for example, in random networks. They are therefore rather lattice--like, so we would expect to observe a relatively high cover time. 

\subsection{Effective medium approximation}

In order to understand the origin of the minima observed through numerical simulations in the pair--averaged FPT and in the mean cover time, we will search here for these features using an analytical approximation. 

In the context of random walks on networks, an \textit{effective medium approximation} (EMA) \cite{bruggeman_berechnung_1935,thiel_effective-medium_2016,sood_first-passage_2005} can be used to solve diffusion problems by approximating the network topology by an ``average structure''. Here, this is done by connecting all pairs of nodes, $u$ and $v$, with a link with weight $p_{vu}$, equal to the probability of connecting both nodes in the original random network.
In order to do this, $p_{vu}$ has to be normalized so that, from each node, the total probability to jump to any other node in one time step is equal to one, that is, $\sum_{v}p_{vu}=1$.
For example, given an Erdős--Rényi network with $N$ nodes, this EMA would connect each node to every other node, except for itself, through a link with weight $p_{vu}=p=1/(N-1)$ and hence the mean FPT between any pair of nodes would be equal to the mean FPT in a complete graph $\tau_{vu} = N-1$, a result consistent with other EMAs \cite{sood_first-passage_2005}.

To develop this EMA, we first build an approximate ``effective'' network where every node is connected to every other node through links with weights given by $p_{\ell}=\tilde{p}_{\ell}/\left\langle k\right\rangle .$
Here, $\tilde{p}_{\ell}$ denotes the probability for two nodes to be connected in the original SSMH model, as in Equation (\ref{eq:p_l}).
On this network, we can then investigate a discrete--time random walk with a sink at an arbitrary node $v$. When a walker is positioned at a node $u$, the probability to jump to a node $v$ that is at hierarchical distance $d(u,v)\equiv d$ is hence
\[
p_{d}=\frac{1}{B-1}\left(\frac{1-\xi}{1-\xi^{L}}\right)\left(\frac{\xi}{B}\right)^{d-1}.
\]
As we derive in Appendix \ref{app:EMA_Random_walk}, the transition matrix from a walker being at hierarchical distance $d>0$ to being at hierarchical distance $d'$ from the sink is given by
\begin{align*}
\tilde{P}{}_{d'd} & =\begin{cases}
\frac{B^{d'-1}}{B^{d-1}}\xi^{d-1}\frac{1-\xi}{1-\xi^{L}}, & d'<d\\
\frac{1-\xi^{d-1}}{1-\xi^{L}}+\frac{1-\xi}{1-\xi^{L}}\xi^{d-1}\frac{B-2}{B-1}, & d'=d\\
\frac{1-\xi}{1-\xi^{L}}\xi^{d'-1}, & d'>d.
\end{cases}
\end{align*}
The pair--averaged FPT given by an EMA is then found to be
\[
\tau_{L}^{\mathrm{E}}=\mathbf{1}^{T}\left[\identity-\tilde{\mathbf{P}}\right]^{-1}\tilde{\mathbf{b}}.
\]
Here, $\mathbf{1}^{T}=(1,1,...,1)$ and the vector $\tilde{\mathbf{b}}$ contains the ratio of all possible sources at layer $d$, with components given by
\[
\tilde{b}_{d}=\frac{B-1}{B^{L}-1}B^{d-1},\qquad1\leq d\leq L.
\]
For example, if we consider $L=2$ hierarchical layers we have 
\[
\tau_{2}^{\mathrm{E}}=\frac{(B-1)B(\xi+1)\left(B^{2}\xi+B\xi+B-\xi\right)}{(B+1)\xi(B\xi+B-\xi)}.
\]
The results for $1\leq L\leq4$ are explicitly given in Appendix \ref{app:EMA_Random_walk}.

\begin{figure*}
\begin{centering}
\includegraphics[width=17cm]{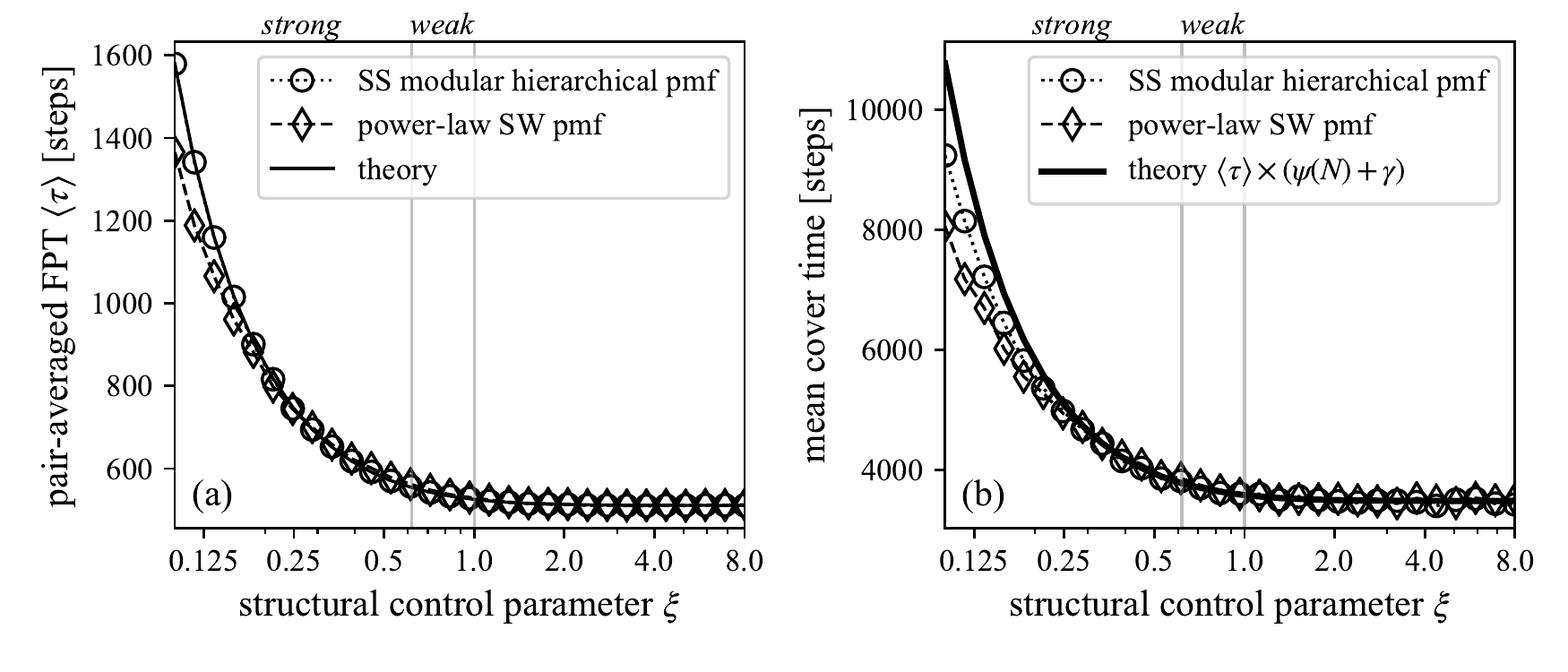}
\par\end{centering}
\raggedright{}\caption{\label{fig:EMA} 
Effective medium approximation results for the pair--averaged first passage time \textbf{(a)} and the mean cover time \textbf{(b)} on SSMH and PLSW networks, as a function of the structural control parameter $\xi$. The circles and diamonds display numerical effective medium simulation results and the dashed lines interpolate between them. Solid lines show the corresponding approximate theoretical results, obtained from Equation (\ref{eq:mean_cover_time}) and from Equation (\ref{eq:GMFPT_L_3}) in Appendix \ref{app:EMA_Random_walk}. 
Parameters are $B=8$ and $L=3$ for all presented results. For the PLSW curves, which depend on the mean degree, we also set $\expv k=7$.}
\end{figure*}

If we assume that the random walk relaxes quickly \cite{maier_cover_2017} (i.e., that it approaches the equilibrium distribution in a small number of steps $t\ll N$), it is furthermore possible to find the mean cover time of the random walk analytically as
\begin{equation}
T_{L}^{\mathrm{E}}=\tau_{L}^{\mathrm{E}}\left(\gamma+\psi(B^{L})\right). \label{eq:mean_cover_time}
\end{equation}
Here, $\gamma$ is the Euler-Mascheroni constant and $\psi(N)=\Gamma'(N)/\Gamma(N)$,
where $\Gamma(N)$ is the Gamma function. We expect this assumption
to lose validity for small $\xi$, where the network becomes more
lattice--like and hence cannot have a quick relaxation time.
Our simulations show that this is indeed the case (see \fig{\ref{fig:EMA}}).

We observe that, when using an EMA, both the pair--averaged FPT and the mean cover time are monotonically decreasing functions of the structural control parameter $\xi$ (or $\mu$, respectively). This would imply that the effectiveness of diffusion processes must increase monotonically as the level of homogeneous randomness in the system is increased. However, this contradicts our numerical results, which show minimal diffusion times at intermediate levels of hierarchical modularity (see Figure~\ref{fig:SSMH_MFPT}). 
The fact that the EMA does not capture this minimum thus leads us to conclude that the detailed topological structure of SSMH and PLSW networks must play a crucial role in generating this effect.

\subsection{Heuristic approach}

\begin{figure*}
    \begin{centering}
    \subfloat{\includegraphics[width=9cm]{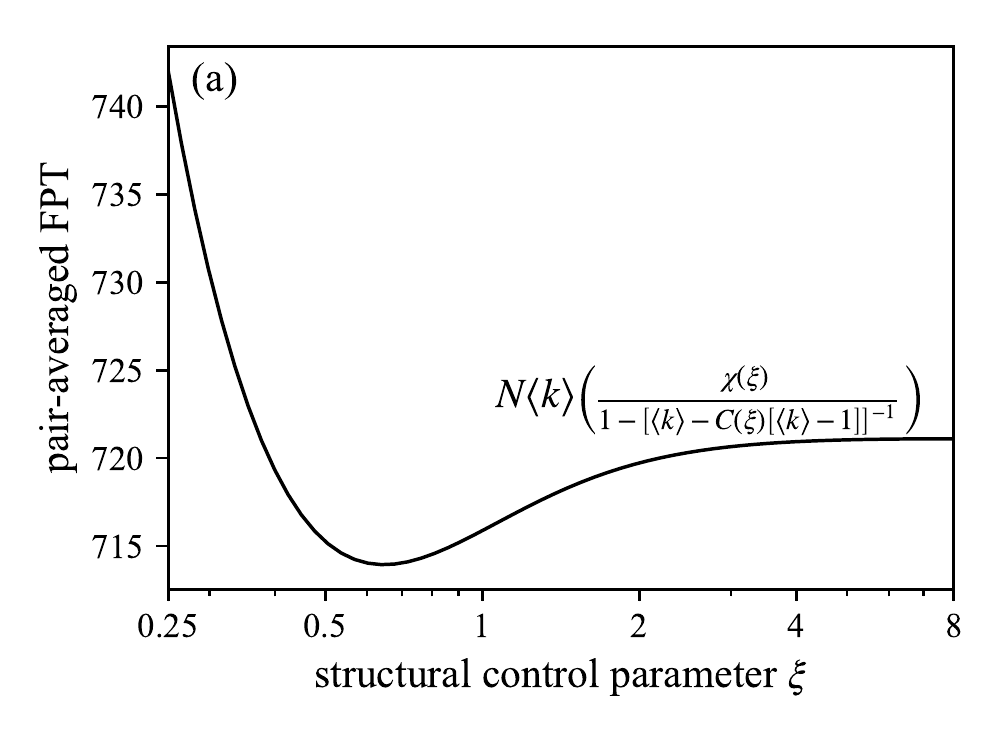}}\subfloat{\includegraphics[width=9cm]{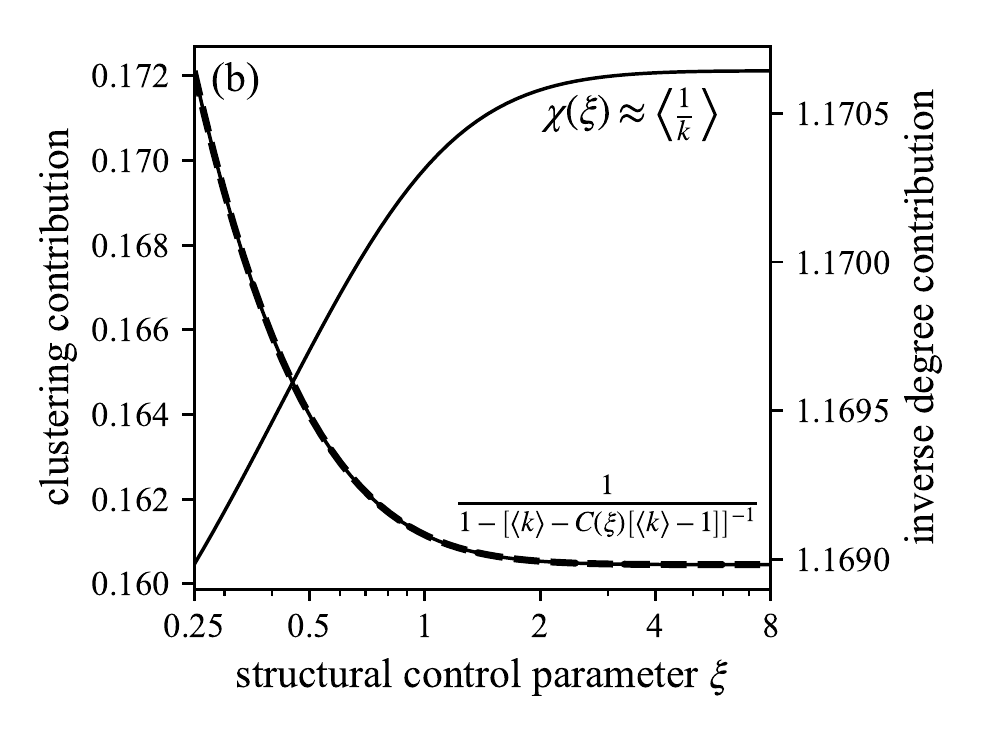}}
    \par\end{centering}
    \raggedright{}\caption{\label{fig:heuristic_MFPT}
     Heuristic result \textbf{(a)} for  the pair--averaged FPT as a function of the structural control parameter $\xi$, as given by Equation (\ref{eq:heuristic_clustered_MGMFPT}) with parameters $B=8$, $L=3$ and $\expv k=7$, and 
    \textbf{(b)} the mean inverse degree and clustering coefficient contributions to this curve, showing how the combination of these growing and decaying functions can result in the observed minimum.}
\end{figure*}

Given that the EMA approach described above fails to reproduce the minimal diffusion time observed numerically at intermediate levels of hierarchical modularity (see Figure~\ref{fig:SSMH_MFPT}), we present here a heuristic approach that will help us understand its origins.
We begin by considering a recent study that found that the global mean FPT of a target node $v$ has a lower bound for networks with short relaxation times and no degree correlation \cite{lau_asymptotic_2010}. This calculation is based on the assumptions that the network is locally tree--like (i.e., that it has vanishing clustering coefficient) and that the mean FPT for any target node $v$ is exponentially distributed as $\exp(-t\beta_{v})$, with
\[
\beta_{v}=\frac{k_{v}}{N\expv k}\left(1-\frac{1}{\expv k}\right).
\]
Since the mean FPT is only asymptotically exponentially distributed, by using this assumption we will obtain a lower bound for the corresponding pair--averaged FPT.
We thus have $\expv{\tau}\geq N^{-1}\sum_v\beta^{-1}_v$, which yields
\[
\expv\tau\geq N\expv k\expv{\frac{1}{k}}_{k>0}\frac{1}{1-\expv k^{-1}},
\]
where $\expv{\cdot}_{k>0}$ denotes the average over all nodes with non--zero degree.
This approximation is not fully valid in our case, however, since the small clustering coefficient assumption becomes more and more incorrect as the level of hierarchical clustering is increased. 
Instead, as shown in Appendix \ref{app:derivation_MGMFPT}, a corrected expression for the pair--averaged FPT that does not need to assume vanishing local clustering can be found through a heuristic approach. In this approach, we replace the neighbors of each focal node $v$ by nodes with degree $\expv k$ and assume that the number of edges between these neighbors is $(1/2) C \expv k(\expv k-1)$, where $C$ is the clustering coefficient. Using this approximation, the derivation detailed in Appendix \ref{app:derivation_MGMFPT} leads to a corrected approximate lower bound for the pair--averaged FPT, which is found to be
\begin{equation}
\expv{\tilde{\tau}}=\expv{\frac{1}{k}}_{k>0}\frac{N\expv k}{1-\Big[\expv k-C\big[\expv k-1\big]\Big]^{-1}}.\label{eq:heuristic_clustered_MGMFPT}
\end{equation}
This function is plotted in Figure~\ref{fig:heuristic_MFPT}a. Its shape as a function of $\xi$ results from two opposing effects (see Figure~\ref{fig:heuristic_MFPT}b): the growth of the mean inverse degree and the decay of the clustering coefficient (as derived in Appendix \ref{app:clustering}) for increasing $\xi$.
The mean inverse degree can be approximated by $\chi(\xi)\propto\expv{k^2}$, which grows monotonically with $\xi$, as shown in Appendix \ref{app:degree_variance}. This results from the fact that the shape of the degree distribution is rather narrow for small $\xi$ values, which leads to a vanishing number of nodes with small degrees. At the same time, this monotonic growth is countered by the clustering behavior because, as the structural control parameter $\xi$ is increased, the network loses its strong clustering and thus the clustering contribution decreases. 
The combination of these two curves leads to the appearance of a minimum in the pair--averaged FPT at intermediate levels of hierarchical modularity, as shown in Figure~\ref{fig:heuristic_MFPT}a.
We note, however, that the position of the minima given by Equation (\ref{eq:heuristic_clustered_MGMFPT}) differs from those in \fig{\ref{fig:SSMH_MFPT}}, due to the heuristic nature of this equation.
%

Finally, our heuristic argument above for the presence of a minimum pair--averaged FPT as a function of $\xi$ would imply that this effect should be observed in a broader range of systems, if the mean inverse degree grows while the clustering coefficient decays. To test this, we also computed the pair--averaged FPT for a variant of the original Watts--Strogatz small--world model and present these results in Appendix~\ref{app:mapping_to_WS_SW}. We thus verified that a minimum pair--averaged FPT persists, even in this case that does not have the modular hierarchical organization or the power--law connection probability that characterize, respectively, our SSMH and PLSW network models.


\section{Discussion and Conclusions}
\label{sec:discussion}

In this paper we studied how modular hierarchical (MH) network structures affect the characteristic timescales of the dynamics of diffusive processes that evolve on these networks. To this end, we first defined the SSMH network class, which interpolates between strongly hierarchically clustered networks and Erdős--Rényi random networks with a single structural control parameter, while keeping the mean degree constant. These networks are similar to those introduced in \cite{newman_modularity_2006}, but were designed to be self--similar at all scales to facilitate our analytical understanding.
Given that previous studies had shown that MH networks can display small--world properties for general search processes, we also defined a PLSW network class (based on a modified Kleinberg small--world model \cite{kleinberg_small--world_2000}) where nodes are embedded in one-dimensional space and are connected with a probability that decays as a power law with distance. This class interpolates between lattice--like nearest neighbor networks and Erdős--Rényi random networks, also using a single structural control parameter, i.e. the power law exponent, while keeping a fixed mean degree. 
We then identified similarities between the topological structures of both models. By considering how the probability of non--local links changes as a function of each structural control parameter, we found a natural way to relate both parameters through an analytical expression.

Next, we compared the dynamics of diffusive processes on the SSMH and PLSW classes of network models, showing that they present similar features. 
We focused, in particular, on the pair--averaged FPT and the mean cover time, two quantities that characterize the typical efficiency of random signaling or distribution processes. 
One of our main results was to observe, through direct numerical simulations, the presence of a minimum for the pair--averaged FPT and the mean cover time on both classes of networks, as a function of their structural control parameter. 
We showed that these optimal diffusion properties are also related to the small--world effects associated to MH structures, appearing to have the same origin as the minimal search time that had been previously observed for targeted localization algorithms in, both, MH \cite{watts_identity_2002} and small--world \cite{kleinberg_small--world_2000} networks. 
This is an interesting extension of previous results, since it does not only apply to targeted searches but also to random search processes that are known to play a prominent role in biological, ecological, and technological systems. 
In the SSMH case, these optimal diffusive properties occur at intermediate levels of hierarchical modularity, where the hierarchy of modules is well--established but there are still enough connections between modules to allow for an effective diffusion between modules. 

In order to explore the origin of the minimal diffusion times observed at intermediate levels of the structural control parameter, we first considered an EMA approach \cite{newman_networks:_2010}, finding that in this approximation the minima are not observed. This implies that the emergence of these minima is a network effect that depends on the specific topological structure, and therefore cannot be explained by replacing the network with an averaged structure. It also implies that, in future investigations, a simple analysis of the average structure of MH network models might not suffice to predict the behavior of dynamical processes on these types of systems. We then provided, instead, a heuristic explanation showing why such minima can be expected for any network where the clustering coefficient decreases while the degree variance and hence the mean inverse degree increases, as a function of a structural control parameter.

In sum, we have sought to understand how the degree of hierarchical modularity of a system may affect its random diffusion properties, in order to explore a potential reason for modular hierarchical structures to be so common in self--organized evolving systems. 
Although we found that MH systems are not only beneficial for targeted search processes, but also for random processes, these properties are not their only relevant features for a living system.  
In future studies, we will thus analyze the effect of MH structures on other classes of dynamic processes, following comparison approaches similar to those presented here. For example, since the pair--averaged FPT is linked to the smallest eigenvalues of the unnormalized graph Laplacian, we can expect MH optimal structures not only for diffusion processes, but similarly for the synchronization of coupled oscillators. By extending our results to other dynamical processes, we will thus achieve a better understanding of the origins and role of MH structures in nature.


\section*{Acknowledgment}

This work was supported by the ``CONNECT-Programm der Alexander von Humboldt-Stiftung''.
D. B. would like to thank Imbish Mortimer for inspiring discussions on the topic.

\appendix

\setcounter{equation}{0}
\renewcommand\theequation{A.\arabic{equation}}
\setcounter{figure}{0}
\renewcommand\thefigure{A.\arabic{figure}}

\section{Relating the structural control parameters of different network models}

\label{sec:mapping_mhrn_to_kleinberg}

We will find here how the structural control parameters, $\xi$ and $\mu$, of the two models discussed in the main text can be related to each other. These two models are the self--similar modular hierarchical (SSMH) network model and the power--law small (PLSW) world network model, which corresponds to a modified small--world Kleinberg model.

Let us first approximate the hierarchical distance from a 
focal node at layer $\ell=0$ as a continuous function.
We begin by computing the connection probability
in layer space, which is given by 
\begin{equation}
p(\ell)=\frac{\xi^{\ell}}{B^{\ell}}p_{0}.\label{eq:layer_connection_density}
\end{equation}
The number of the focal node's potential neighbors in layer space
is thus given as 
\[
dn=\frac{dn}{d\ell}d\ell=B^{\ell}\log B\ d\ell,
\]
where we used the number of reachable nodes $n(\ell)=B^{\ell}.$ 
The number of connections the focal node has to layer $\ell$ is therefore
\begin{align*}
dk & =p(\ell)\ dn=\frac{\xi^{\ell}}{B^{\ell}}B^{\ell}p_{0}\log B\ d\ell\\
 & =\xi^{\ell}p_{0}\log B\ d\ell.
\end{align*}
The connection density (connections per layer) in layer space is defined
as 
\[
dk=\frac{dk}{d\ell}d\ell,
\]
so we obtain 
\begin{align*}
\frac{dk}{d\ell} & =\xi^{\ell}p_{0}\log B\equiv c(\ell).
\end{align*}
Now, with $p(y)=\frac{dy}{dx}p(x)$ and $\ell(n)=\log n/\log B$ we
find
\begin{align*}
c(n) & =\frac{1}{n\log B}\xi^{\ell(n)}p_{0}\log B=p_{0}n{}^{\frac{\log\xi}{\log B}-1}\\
 & =p_{0}n{}^{\mu-1}
\end{align*}
with $\mu=\log\xi/\log B$ which is Kleinberg's distance--based power--law
connection probability in a one--dimensional lattice.
By considering periodic boundary conditions and treating the focal node as centered
with symmetric connection probability to the left and right, s.t.
two nodes can only have lattice distance $-N/2\leq n\leq N/2$, and so
\[
c_{n}=p_{0}|n|^{\mu-1}.
\]
We require here again that the mean degree is fixed, so the normalizing
constant evaluates to 
\begin{equation}
p_{0}=\expv k\left(2\sum_{n=1}^{\left\lfloor N/2\right\rfloor }n^{\mu-1}+\mathrm{mod}(N,2)(\left\lfloor N/2\right\rfloor +1)^{\mu-1}\right)^{-1},\label{eq:kleinberg_norm}
\end{equation}
where $\mathrm{mod}(N,2)$ denotes the remainder of the integer division $N/2$
and $\left\lfloor \cdot\right\rfloor $ is the floor function. 

\begin{figure}
    \centering
    \includegraphics[width=3.375in]{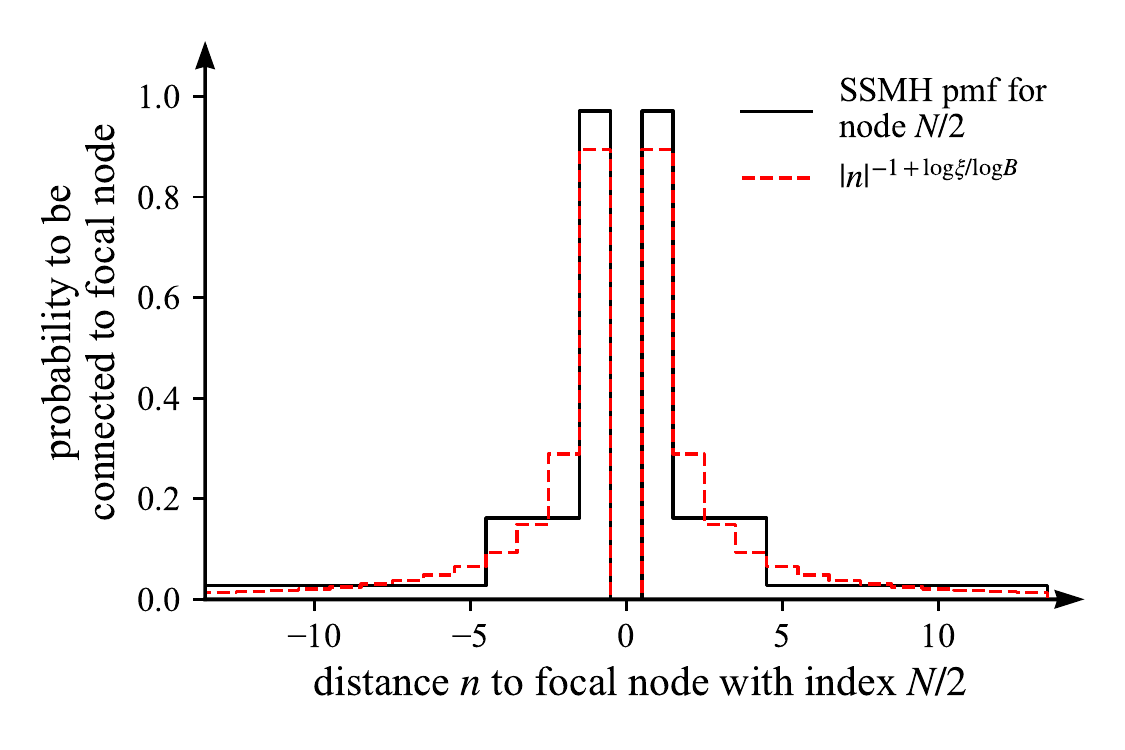}\,\caption{\label{fig:SSMH_pmf}
    Comparison of the probability mass functions associated with SSMH networks and with standard one--dimensional Kleinberg networks.}
\end{figure}

We can interpret these results as showing that the SSMH connection probability mass function corresponds approximately to a discretized version of the power--law $c_{n}$ that is associated to the PLSW connection probability, but where target nodes are grouped in exponentially growing batches (cf. Figure \ref{fig:SSMH_pmf}). In this approximation, however, the additional embedded nature of the SSMH network structure is lost.

\setcounter{equation}{0}
\renewcommand\theequation{B.\arabic{equation}}
\setcounter{figure}{0}
\renewcommand\thefigure{B.\arabic{figure}}

\section{Effective medium approximation for random walks on SSMH networks}
\label{app:EMA_Random_walk}
In what follows, we show how to compute FPT statistics for structurally averaged SSMH networks in an \textit{effective medium approximation} (EMA), in which the possible edges are replaced by the probability for those edges to exist.

Following Equation (\ref{eq:p_l}), we
denote the probability of two nodes to be connected as $\tilde{p}_{\ell}$.
In the context of this EMA, this means that every node is connected to every other node
but edges are weighted with probability $p_{\ell}=\tilde{p}_{\ell}/\left\langle k\right\rangle .$
On this network, we investigate a random walk with a sink at an arbitrary
node $v$. When a walker is positioned at node $u$, the probability
to jump to node $v$ at hierarchical distance $d(u,v)\equiv d$ is
\[
p_{d}=\frac{1}{B-1}\left(\frac{1-\xi}{1-\xi^{L}}\right)\left(\frac{\xi}{B}\right)^{d-1},
\]
as reasoned above. 
In this context, the following scenarios may happen.
\begin{enumerate}
\item The walker jumps to node $v$. This will happen with probability 
\[
p_{1}=p_{d}.
\]
The new distance to node $v$ will be $d'=0$.
\item The walker jumps to a node in a lower layer $\ell<d$. The new distance
will still be $d'=d$. In each layer $\ell<d$, there will be $B^{\ell-1}(B-1)$
possible target nodes for the walker starting at node $u$. The probability
for this scenario to happen is
\[
p_{2}=\sum_{\ell=1}^{d-1}p_{\ell}B^{\ell-1}(B-1)=\frac{1-\xi^{d-1}}{1-\xi^{L}}.
\]
\item The walker jumps to a node in $\ell=d$, but misses the branch that
$v$ is in and the branch that $u$ is in (because this event is covered
in scenario 2). The new distance is still $d'=d$. There are $B^{d-1}(B-2)$
nodes for this event. Consequently, it happens with probability
\[
p_{3}=p_{d}B^{d-1}(B-2).
\]
\item The walker jumps to a node in $\ell=d$, hits the branch that $v$
is in, but misses $v$. The new distance is $d'<d$, for each $d'$
with probability
\[
p_{4}=p_{d}B^{d'-1}(B-1).
\]
\item The walker jumps to a node in $\ell>d$. Consequently, the new distance
is $d'>d$. The probability of this happening is, for each $d'$
\[
p_{5}=p_{d'}B^{d'-1}(B-1).
\]
\end{enumerate}
The transition matrix from a walker being at distance $d$ from the
target to being at new distance $0\leq d'\leq L$ from the target
is hence
\begin{align*}
P_{d'd} & =\begin{cases}
\frac{1}{(B-1)}\frac{1-\xi}{1-\xi^{L}}\left(\frac{\xi}{B}\right)^{d-1}, & d'=0\ \wedge\ d>0\\
\frac{B^{d'-1}}{B^{d-1}}\xi^{d-1}\frac{1-\xi}{1-\xi^{L}}, & d'<d\ \wedge\ d>0\\
\frac{1-\xi^{d-1}}{1-\xi^{L}}+\frac{1-\xi}{1-\xi^{L}}\xi^{d-1}\frac{B-2}{B-1}, & d'=d\ \wedge\ d>0\\
\frac{1-\xi}{1-\xi^{L}}\xi^{d'-1}, & d'>d\ \wedge\ d>0\\
0, & d=0\ \wedge\ d=0\\
1, & d'=0\ \wedge\ d=0.
\end{cases}
\end{align*}
We define the vector $\mathbf{p}^{(d_{\mathrm{start}})}(t)$ as the
probability to find a single walker at distance $d$ when a random
walker started at distance $d_{\mathrm{start}}$ from the target.
Hence, the vector has $L+1$ entries, ranging between $0$ and $L$.
Let a discrete--time random walk begin with initial conditions given
by the vector 
\[
p_{d}^{(d_{\mathrm{start}})}(0)=\delta_{dd_{\mathrm{start}}},
\]
i.e. there is a single walker at distance $d_{\mathrm{\mathrm{start}}}$.
The probability to find the random walker at distance $d$ when starting
at distance $d_{\mathrm{start}}$ after $t$ time steps is
\begin{equation}
p_{d}^{(d_{\mathrm{start}})}(t)=\left(\mathbf{P}^{t}\right)_{d,d_{\mathrm{start}}}.\nonumber
\end{equation}
Suppose, however, that we start with one random walker on every node $u\neq v$.
Then the total ratio of walkers that are absorbed up to a time $t$
is 
\begin{align}
p_{0}(t) & =\frac{1}{B^{L}-1}\sum_{d_{s}=1}^{L}B^{d_{s}-1}(B-1)p_{0}^{(d_{s})}(t)\label{eq:all_starter_contributions}\\
 & =\frac{1}{B^{L}-1}\sum_{d_{s}=1}^{L}B^{d_{s}-1}(B-1)\left(\mathbf{P}^{t}\right)_{0,d_{s}}.\nonumber 
\end{align}
Another way to write down the ratio of walkers absorbed into the sink
node at time $t$ is 
\[
p_{0}^{(d_{s})}(t)=1-\sum_{d=1}^{L}p_{d}^{(d_{s})}(t),
\]
since the probability of the walker being at any distance $0\leq d\leq L$
is equal to one and conserved at all times. We continue by introducing a
few new quantities. As described above, the probability of being at
distance $d$' at time $t$ is
\[
p_{d'}(t)=\sum_{d=0}^{L}P_{d'd}p_{d}(t-1).
\]
However, for $d'>0$, the column $d=0$ does not contribute anything
to the sum (since it is filled with zeros). Hence, we define a new
vector $\tilde{\mathbf{p}}^{(d_{\mathrm{start}})}$ which is the former
probability vector with the $0$-th element removed and a new transition
matrix $\tilde{\mathbf{P}}$, as the old transition matrix with the
$0$-th column and row removed.
We therefore have
\[
\tilde{p}_{d}^{(d_{\mathrm{start}})}(t)=\left(\tilde{\mathbf{P}}^{t}\right)_{d,d_{\mathrm{start}}},
\]
such that
\[
p_{0}^{(d_{s})}(t)=1-\sum_{d=1}^{L}\left(\tilde{\mathbf{P}}^{t}\right)_{d,d_{s}}.
\]
In order to consider the contribution of all starting nodes, we combine
these results with Equation (\ref{eq:all_starter_contributions})
to find
\[
p_{0}(t)=1-\frac{B-1}{B^{L}-1}\sum_{d_{s}=1}^{L}B^{d_{s}-1}\sum_{d=1}^{L}\left(\tilde{\mathbf{P}}^{t}\right)_{d,d_{s}}.
\]
This is the ratio of walkers that have been absorbed in to the sink
$v$ up until time $t$. We further introduce the vector $\tilde{\mathbf{b}}$
containing the fractions of all possible targets at layer $d$ as
\[
\tilde{b}_{d}=\frac{B-1}{B^{L}-1}B^{d-1},\qquad1\leq d\leq L,
\]
as well as the vector $\mathbf{1}^{T}=(1,1,...,1)$. 
Our observable then reduces to
\begin{equation}
p_{0}(t)=1-\mathbf{1}^{T}\tilde{\mathbf{P}}^{t}\tilde{\mathbf{b}}.\label{eq:cdf_arrival_time}
\end{equation}
We now focus on finding the global mean FPT for
the focal node. This quantity can be calculated as 
\begin{align*}
\tau & =\sum_{t=0}^{\infty}t\big[p_{0}(t)-p_{0}(t-1)\big]=\sum_{t=0}^{\infty}t\left[\mathbf{1}^{T}\tilde{\mathbf{P}}^{t-1}\tilde{\mathbf{b}}-\mathbf{1}^{T}\tilde{\mathbf{P}}^{t}\tilde{\mathbf{b}}\right]\\
 & =\mathbf{1}^{T}\left[\identity-\tilde{\mathbf{P}}\right]^{-1}\tilde{\mathbf{b}}.
\end{align*}
Note that this quantity is equal to, both, the global mean FPT and the pair--averaged FPT (that is, the mean global mean FPT), because in this EMA all nodes are equal. This result is similar
to the result for arbitrary networks, where instead of $(\identity-\tilde{\mathbf{P}})^{-1}$
one uses the inverse of the reduced unnormalized graph Laplacian.
However, using the layer approach, we can reduce the matrix size from
$B^{L}-1$ to $L$, a significant reduction in degrees of freedom.
This makes it possible to obtain analytical expressions
for the global mean FPT. For examples, for $L<5$ we have
\begin{widetext}
\begin{align}
L & =1,\qquad\tau=B-1\nonumber \\
L & =2,\qquad\tau=\frac{(B-1)B(\xi+1)\left(B^{2}\xi+B\xi+B-\xi\right)}{(B+1)\xi(B\xi+B-\xi)}\nonumber \\
L & =3,\qquad\tau=\frac{(B-1)B^{2}\left(\xi^{2}+\xi+1\right)\left(B^{4}\xi^{2}(\xi+1)+B^{3}\xi(\xi+1)+B^{2}(\xi+1)^{2}-B\xi\left(2\xi^{2}+3\xi+2\right)+\xi^{2}(\xi+1)\right)}{\left(B^{2}+B+1\right)\xi^{2}(B\xi+B-\xi)\left(B\left(\xi^{2}+\xi+1\right)-\xi(\xi+1)\right)}\label{eq:GMFPT_L_3}\\
L & =4,\qquad\tau=(B-1)B^{3}(\xi+1)\left(\xi^{2}+1\right)\bigg[B^{6}\xi^{3}\left(\xi^{3}+2\xi^{2}+2\xi+1\right)-B^{5}\xi^{2}\left(\xi^{4}+\xi^{3}-2\xi-1\right)+\nonumber \\
 & \qquad\qquad\qquad+B^{4}\xi(\xi+1)^{2}+B^{3}\left(\xi^{4}+2\xi^{3}+3\xi^{2}+3\xi+1\right) -B^{2}\xi\left(2\xi^{5}+6\xi^{4}+10\xi^{3}+11\xi^{2}+8\xi+3\right)+\nonumber \\
 & \qquad\qquad\qquad+B\xi^{2}\left(3\xi^{4}+7\xi^{3}+9\xi^{2}+7\xi+3\right)-\xi^{3}\left(\xi^{3}+2\xi^{2}+2\xi+1\right)\bigg]\nonumber \\
 & \qquad\qquad\qquad\Bigg/\ \bigg[\left(B^{3}+B^{2}+B+1\right)\xi^{3}(B\xi+B-\xi)\left(B\left(\xi^{2}+\xi+1\right)-\xi(\xi+1)\right)\times\nonumber\\
  & \qquad\qquad\qquad\qquad\times\left(B\left(\xi^{3}+\xi^{2}+\xi+1\right)-\xi\left(\xi^{2}+\xi+1\right)\right)\bigg].\nonumber 
\end{align}
\end{widetext}

We note that for $\xi\rightarrow B$
the global mean FPT approaches the Erdős--Rényi solution $\tau=B^{L}-1$. 
In addition, the global mean FPT diverges for $\xi\rightarrow0$, since the effective
medium then approaches a state where it consists of $B^{L-1}$ complete
networks, each containing $B$ nodes. 

\setcounter{equation}{0}
\renewcommand\theequation{C.\arabic{equation}}
\setcounter{figure}{0}
\renewcommand\thefigure{C.\arabic{figure}}

\section{Properties of self--similar modular hierarchical networks}
\label{app:SSMH_properties}

We derive in this Appendix, three important quantities that characterize self--similar modular hierarchical (SSMH) networks as a function of the structural control parameter $\xi$: the degree variance, the moment generating function, and the clustering coefficient as quantified by the transitivity.

\subsection{Degree variance}
\label{app:degree_variance}

The variance of the degree can be an important control parameter for certain dynamic processes on networks. It controls, for example, the transition to an epidemic spreading processes in complex networks \cite{barrat_dynamical_2008}. We derive here the degree variance for SSMH networks, as a function of the mean degree $\expv{k}$, number of modules per level $B$, number of levels $L$, and structural control parameter $\xi$.

We begin by computing the second moment of the degree distribution, which is given by

\begin{align*}
\expv{k^{2}} & =\expv{\left(\sum_{\ell=1}^{L}k_{\ell}\right)^{2}}\\
 & =\sum_{\ell=1}^{L}\expv{k_{\ell}^{2}}+2\sum_{\ell=2}^{L}\sum_{m=1}^{\ell-1}\underbrace{\expv{k_{\ell}k_{m}}}_{=\expv{k_{\ell}}\expv{k_{m}}}.
\end{align*}
Since the layer degrees are binomially distributed, the moments of each hierarchical layer are
\begin{align*}
\expv{k_{\ell}} & =B^{\ell-1}(B-1)p_{\ell}\\
\expv{k_{\ell}^{2}} & =B^{\ell-1}(B-1)p_{\ell}\left(1+B^{\ell-1}(B-1)p_{\ell}-p_{\ell}\right).
\end{align*}
We therefore have
\begin{align*}
\sum_{\ell=1}^{L}\expv{k_{\ell}^{2}} & =\expv{k}+\expv{k}^{2}\frac{(1-\xi)(1+\xi^{L})}{(1+\xi)(1-\xi^{L})}-\\
 & \ \ \ -\frac{\expv{k}^{2}}{B-1}\left(\frac{1-\xi}{1-\xi^{L}}\right)^{2}\left(\frac{1-(\xi^{2}/B)^{L}}{1-\xi^{2}/B}\right)
\end{align*}
and 
\begin{align*}
\sum_{\ell=2}^{L}\sum_{m=1}^{\ell-1}\expv{k_{l}}\expv{k_{m}} & =\expv{k}^{2}\left[\frac{1}{1-\xi^{L}}-\frac{1-\xi^{2L}}{(1-\xi^{L})^{2}}\frac{1}{1+\xi}\right].
\end{align*}
By combining these results, we obtain
\begin{align}
\expv{k^{2}} & =\expv{k}+\expv{k}^{2}\left[1-\frac{1}{B-1}\left(\frac{1-\xi}{1-\xi^{L}}\right)^{2}\left(\frac{1-(\xi^{2}/B)^{L}}{1-\xi^{2}/B}\right)\right],\label{eq:second_moment}
\end{align}
from where the variance of the degree is found to be
\[
\mathrm{Var}(k)=\expv{k}-\frac{\expv{k}^{2}}{B-1}\left(\frac{1-\xi}{1-\xi^{L}}\right)^{2}\left(\frac{1-(\xi^{2}/B)^{L}}{1-\xi^{2}/B}\right).
\]

\subsection{Moment generating function}

Higher moments of the SSMH network degree distribution (beyond the mean and variance computed above) can be helpful to further analyze dynamical processes on these systems.  We thus derive here the associated moment generating function, from which all the SSMH network degree distribution moments can be easily computed. 

The mean degree of a node in layer $\ell$ is binomially distributed. The corresponding parameters are the number of trials $k_{\ell}^{\max}$ and the probability $p_{\ell}$, given by Equations (\ref{eq:k_l_max}) and (\ref{eq:p_l}), respectively.
The moment generating function of layer $\ell$ must thus be given by the generating function of a binomial distribution with the form
\begin{align*}
G_{k_{\ell}}(x) & =\left[1-(1-x)p_{\ell}\right]^{B^{\ell-1}(B-1)}\\
 & =\left[1-(1-x)\frac{\left\langle k\right\rangle }{B-1}\left(\frac{1-\xi}{1-\xi^{L}}\right)\left(\frac{\xi}{B}\right)^{\ell-1}\right]^{B^{\ell-1}(B-1)}.
\end{align*}
Since the combined degree distribution of all layers $k$ is given by the convolution of all $k_{\ell}$, the moment generating function is found to be
\begin{align*}
G_{k}(x) & =\prod_{\ell=1}^{L}G_{k_{\ell}}(x)\\
 & =\prod_{\ell=1}^{L}\left[1-(1-x)\frac{\left\langle k\right\rangle }{B-1}\left(\frac{1-\xi}{1-\xi^{L}}\right)\left(\frac{\xi}{B}\right)^{\ell-1}\right]^{B^{\ell-1}(B-1)}.
\end{align*}

\subsection{Clustering}
\label{app:clustering}
We now compute the clustering coefficient as quantified by the transitivity as a function of the structural control parameter for SSMH networks. The clustering coefficient typically directly affects the relaxation of diffusion processes, with highly clustered networks usually displaying a slow relaxation behavior when compared to networks with low clustering. Subsequently, we compare the computed transitivity to the mean local clustering coefficient numerically and find that both follow similar qualitative behavior.

The mean transitivity is defined by the ensemble mean 
\[
    \expv{C}=\expv{\frac{3\bigtriangleup}{3\bigtriangleup+\wedge}}.
\]
Here, $\bigtriangleup$ is the number of unique triangles in a single network and $\wedge$ is the number of non--closed triads. The mean transitivity is thus a measure for the presence of triadic closures.
Given the adjacency matrix $\mathbf{A}$ of a single network realization, the transitivity can be calculated as
\[
C=\mathrm{Tr}\left(\mathbf{A}^{3}\right)\Big/\mathrm{Tr}\left(\mathbf{A}^{2}\right).
\]

In what follows, we approximate the ensemble mean as a mean over triads such that
\[
\expv{C} \approx \frac{\expv{3\bigtriangleup}}{\expv{3\bigtriangleup+\wedge}}.
\]
We replace the adjacency matrix with the matrix $\mathbf{\Pi}$ where each entry $\Pi_{ij}$ is given by
the probability that $i$ and $j$ are connected as defined in Equation~(\ref{eq:p_l}), 
\[ 
\Pi_{ij}=p_{\ell(i,j)}.
\]
Here, $\ell(i,j)$ is the lowest layer in which $i$ and $j$ are part of
the same group. Because all nodes in the network defined by $\mathbf{\Pi}$
can be considered as ``equal'', it suffices to calculate the mean transitivity as 
\[
\expv{C}\approx\left(\mathbf{\Pi}^{3}\right)_{ii}\Big/\left(\mathbf{\Pi}^{2}\right)_{ii}
\]
where $i$ is any node in the network. For simplicity and without
loss of generality we set $i=1$. The numerator is then given by
\[
\left(\mathbf{\Pi}^{3}\right)_{11}=2\sum_{j>1}^{N-1}\sum_{u>j}^{N}p_{\ell(1,j)}p_{\ell(1,u)}p_{\ell(j,u)}.
\]
Hence, for every pair of nodes $(j,u)$ with $j\neq u\neq1$, we calculate
the probability that node 1 is connected to node $j$, node 1 is
connected to node $u$, and node $j$ is connected to node $u$. Analogously,
the denominator is the sum over all pairs $(j,u)$ of the probability
that node $1$ is connected to $j$ and 1 is connected to $u$ (thus
building a triad),
\[
\left(\mathbf{\Pi}^{2}\right)_{11}=2\sum_{j>1}^{N-1}\sum_{u>j}^{N}p_{\ell(1,j)}p_{\ell(1,u)}.
\]
Instead of summing over all nodes, we want to sum over layers, as
this significantly reduces the complexity of the calculation.
We thus split both sums into two contributions
\begin{align*}
\frac{1}{2}\left(\mathbf{\Pi}^{3}\right)_{11} & \equiv\pi^{(3)}=\pi_{S}^{(3)}+\pi_{L}^{(3)}\\
\frac{1}{2}\left(\mathbf{\Pi}^{2}\right)_{11} & \equiv\pi^{(2)}=\pi_{S}^{(2)}+\pi_{L}^{(2)}
\end{align*}
where subscripts represent (S)hort--range and (L)ong--range contributions.
We begin with the evaluation of the long range contributions by considering the following scenario. 
The pair $(1,j)$ has
hierarchical distance $\ell_{1}$, whereas the pair $(1,u)$ has hierarchical
distance $\ell_{1}<\ell_{2}$. This means that nodes 1 and $j$ share
a subgroup in layer $\ell_{1}$ and thus, if the third node $u$ has
distance $\ell_{2}$ to node 1, so has node $j$ distance $\ell_{2}$
to node $u$. Since the problem is symmetrical, the case $\ell_{1}>\ell_{2}$
will contribute the same amount and hence it suffices to look at $\ell_{1}<\ell_{2}$.
Now there's $B^{\ell_{1}}(B-1)$ possible target nodes for node 1
in layer $\ell_{1}$ and $B^{\ell_{2}}(B-1)$ possible target nodes
in layer $\ell_{2}$. Consequently, the total number of potential
pairs in layer combination $\ell_{1},\ell_{2}$ is $B^{\ell_{1}}(B-1)\times B^{\ell_{2}}(B-1)$
and the total sum evaluates to
\begin{align*}
\pi_{L}^{(3)} & =\sum_{\ell_{1}=1}^{L-1}\sum_{\ell_{2}=\ell_{1}+1}^{L}(B-1)^{2}B^{\ell_{1}}B^{\ell_{2}}p_{\ell_{1}}p_{\ell_{2}}p_{\ell_{2}}\\
 & =\frac{\left\langle k\right\rangle ^{3}}{B-1}\left(\frac{1-\xi}{1-\xi^{L}}\right)^{3}\frac{1}{1-\xi^{2}/B}\times\\
 & \qquad\times\left[\left(\frac{\xi^{2}}{B}\right)\frac{1-(\xi^{3}/B)^{L-1}}{1-\xi^{3}/B}-\left(\frac{\xi^{2}}{B}\right)^{L}\left(\frac{1-\xi^{L-1}}{1-\xi}\right)\right].
\end{align*}

The short--range contributions stem from pairs of target nodes $(j,u)$
where node $1$ has the same distance $\ell$ to both of them, hence
they build a triad with probability $p_{\ell}^{2}$ . We make a distinction
between two cases.
\begin{enumerate}
\item Both $j$ and $u$ have distance $\ell$ to node 1, and are part of the same subgroup.
This means that their distance is $\ell'<\ell$ and that they are connected
with probability $p_{\ell'}$. The total number of possible pairs
of distance $\ell'<\ell$ is $(1/2)\times B^{\ell}(B-1)\times B^{\ell'-1}(B-1).$
The additional factor $1/2$ emerges to avoid double counting
(with once $j$ as source and once $u$ as source).
\item Both $j$ and $u$ have distance $\ell$ to node 1, but are not part of the same
subgroup in $\ell'\leq\ell$. This means that $u$ is at distance
$\ell$ of node $j$ but the number of submodules that $j$ can choose
a neighbor from is reduced by two (its own subgroup and the subgroup
of node 1). Hence, the total number of distinct pairs of this type 
is $(1/2)\times B^{\ell-1}(B-2)B^{\ell-1}(B-1).$ There are $B^{\ell-1}(B-1)$
nodes to pick as first neighbor of $1$ and $B^{\ell-1}(B-2)$ nodes
to pick as second neighbor of $1$. Again, there is an additional
factor $1/2$ to avoid double counting.
\end{enumerate}
Considering these cases, we can evaluate the short--range contribution
as follows, (note that we make use of the Kronecker symbol $\delta_{ij}$),
\begin{align*}
\pi_{S}^{(3)} & =\frac{1}{2}\sum_{\ell=1}^{L}p_{\ell}^{2}\sum_{\ell'=1}^{\ell}B^{\ell-1}(B-1)B^{\ell'-1}(B-1-\delta_{\ell\ell'})p_{\ell'}\\
 & =\frac{1}{2}\frac{\left\langle k\right\rangle ^{3}}{(B-1)^{2}}\left(\frac{1-\xi}{1-\xi^{L}}\right)^{3}(B-2)\frac{1-(\xi^{3}/B)^{L}}{1-\xi^{3}/B}+\\
 & \qquad+\frac{B-1}{1-\xi}\left(\frac{1-(\xi^{2}/B)^{L}}{1-\xi^{2}/B}-\frac{1-(\xi^{3}/B)^{L}}{1-\xi^{3}/B}\right).
\end{align*}
The contributions $\pi_{L}^{(2)}$ and $\pi_{S}^{(2)}$ can be calculated
with an analogous procedure as above by setting $p_{\ell_{2}}=1$
and $p_{\ell'}=1$, respectively. We thus find 
\begin{align*}
\pi_{L}^{(2)} & =\frac{\left\langle k\right\rangle ^{2}(1-\xi)}{(1-\xi^{L})^{3}}\left[\xi\frac{1-(\xi^{2})^{L-1}}{1-\xi^{2}}-\frac{\xi^{L}-\xi^{2L-1}}{1-\xi}\right]\\
\pi_{S}^{(2)} & =\frac{1}{2}\frac{\left\langle k\right\rangle ^{2}}{B-1}\left(\frac{1-\xi}{1-\xi^{L}}\right)^{2}\left[(B-1)\frac{1-\xi^{2L}}{1-\xi^{2}}-\frac{1-(\xi^{2}/B)^{L}}{1-\xi^{2}/B}\right].
\end{align*}
\begin{figure}
\begin{centering}
\includegraphics[width=3.375in]{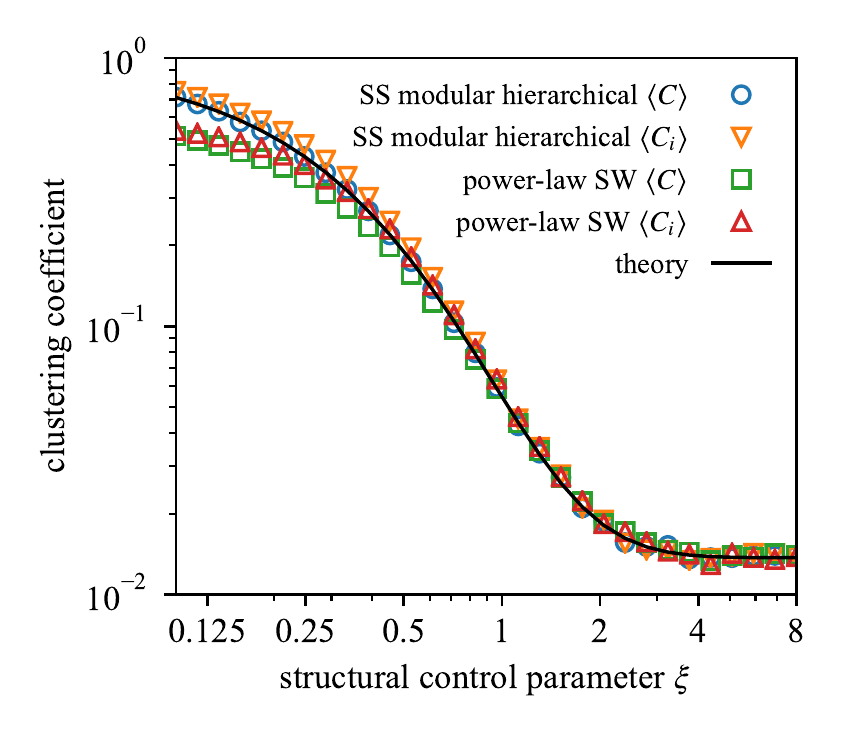}
\par\end{centering}
\raggedright{}\caption{\label{fig:clustering}
Degree of clustering as given by the mean transitivity $\expv{C}$ and the mean local clustering coefficient $\expv{C_{i}}$ for, both, the self--similar modular hierarchical and the power--law small world network models with $B=8$, $L=3$ and $\expv k=7$.
The markers display averages over 20 numerical realizations. 
The solid line shows theoretical prediction of the mean transitivity, given by Equation (\ref{eq:clustering_coefficient}).}
\end{figure}
Hence, the mean transitivity can be approximated as 
\begin{equation}
\expv{C}\approx\frac{\pi_{S}^{(3)}+\pi_{L}^{(3)}}{\pi_{S}^{(2)}+\pi_{L}^{(2)}}.\label{eq:clustering_coefficient}
\end{equation}
As one can see in Figure \ref{fig:clustering}, the derived formula predicts
the SSMH transitivity well and also provides a good approximation for the PLSW network model for intermediate values of the structural control parameter $\xi$ ($\mu$, respectively).

Another clustering measure that is often used is the local clustering coefficient of a node $i$ with degree $k_i$. It is defined as the ratio of existent triangles $\bigtriangleup_i$ of which $i$ is part, compared to the number of potential triangles it could be part of. That is
\[
    C_{i} = \begin{cases}
                0, & k_i \leq 1\\
                2\bigtriangleup_i / (k_i(k_i-1)), & k_i > 1.
            \end{cases}
\]
Figure \ref{fig:clustering} shows that the formula derived in Equation~(\ref{eq:clustering_coefficient}) also approximates the mean local clustering coefficient reasonably well for both the SSMH and PLSW network models.

\setcounter{equation}{0}
\renewcommand\theequation{D.\arabic{equation}}
\setcounter{figure}{0}
\renewcommand\thefigure{D.\arabic{figure}}

\section{Global mean FPT for locally clustered networks}
\label{app:derivation_MGMFPT}
In this Appendix, we show how to compute the global mean FPT for a target sink node in networks that are not locally tree--like.

An effective method to compute the global mean FPT of any target node in a network
was recently found by Lau, \textit{et al.} \cite{lau_asymptotic_2010}. Lau's
approach treats the time dependent walker concentration $P_{u}(t)$
on any node $u$ at time $t$ as composed of two separable distributions 
\[
P_{u}(t)\approx P_{u}P_{\mathrm{total}}(t).
\]
This accounts for a time--independent distribution $P_{u}$ of walker
density on node $u$ and for the total density of walkers $P_{\mathrm{total}}(t)$ 
spread throughout the network. The total walker density
leaving the network to end up in sink $i$ is proportional to the
amount of walkers leaving neighbors of $i$, which we will denote
as $\mathrm{Nei}(i)$. The outflux of walker density at time step
$t\rightarrow t+1$ is hence
\[
\beta_{i}(t)=\sum_{u\in\mathrm{Nei}(i)}\frac{P_{u}(t)}{k_{u}}.
\]
This means that the total walker density in the network is changing as
\begin{align*}
P_{\mathrm{total}}(t+1)-P_{\mathrm{total}}(t) & =-\beta_{i}(t)\\
 & =-P_{\mathrm{total}}(t)\sum_{u\in\mathrm{Nei}(i)}\frac{P_{u}}{k_{u}}\\
 & =-\beta_{i}P_{\mathrm{total}}(t).
\end{align*}
Therefore, by recursion we have
\[
P_{\mathrm{total}}(t)=P_{\mathrm{total}}(0)(1-\beta_{i})^{t}\stackrel{\beta_{i}\ll1}{\approx}P_{\mathrm{total}}(0)\exp(-\beta_{i}t).
\]
In the $P_{\mathrm{total}}(0)=1$ case, we coincidentally find the approximate
cumulative distribution function of the FPT at target
node $i$ as $p(t)=1-\exp(\beta_{i}t)$. Note that the decay rate
is then given by 
\[
\beta_{i}=\sum_{u\in\mathrm{Nei}(i)}\frac{P_{u}}{k_{u}}.
\]
In order to compute the decay rate $\beta_{i}$, and hence the global
mean FPT given by $\tau_{i}=\beta_{i}^{-1}$, we have to estimate the 
walker concentration $P_{u}$ on the neighbors of the sink. Lau \textit{et al.} propose
that the second neighbors $v$ of sink node $i$, with step distance $d(i,v)=2$,
have an approximate equilibrium density given by $P_{v}=P_{v}^{\star}=k_{v}/N\expv k$.
Using this approximation, we find
\begin{equation}
P_{u}=\sum_{v\in\mathrm{Nei}(u)\backslash i}\frac{P_{v}}{k_{v}}=\frac{1}{N\expv k}(k_{u}-1).\label{eq:walker_density_zero_clustering}
\end{equation}
This includes the assumption that all walkers flowing into neighbor
node $u$ originate from second neighbors $v$, which implies that the neighbors
of $i$ are not connected and, hence, that $i$ has local clustering coefficient
$C_{i}=0$. The decay rate of sink node $i$ is thus given by 
\[
\beta_{i}=\frac{1}{N\expv k}\sum_{u\in\mathrm{Nei}(i)}\frac{k_{u}-1}{k_{u}}.
\]
Replacing all neighbors $u$ with $k_{i}$ average neighbors and using
$\expv{k_{\mathrm{neigh}}^{-1}}=\expv{k}^{-1}$, which is valid for uncorrelated networks,
we find 
\[
\beta_{i}=\frac{k_{i}}{N\expv k}\left(1-\frac{1}{\expv k}\right).
\]
This result, however, is based on the assumption of a vanishing clustering coefficient,
which is a crude approximation for SSMH networks in the hierarchically clustered regime.
We therefore obtained a better approximation through the procedure described below.

Consider sink node $i$ with degree $k_{i}$. On average, a neighbor
$u$ of $i$ will have $\kappa_{i}=C_{i}(k_{i}-1)$ neighbors that
are also neighbors of $i$. The rest of its $k_{u}-1-\kappa_{i}$
neighbors each contribute an influx of $1/N\expv k$ walkers, such
that 
\begin{equation}
P_{u}=\frac{1}{N\expv k}\big[k_{u}-1-\kappa_{i}\big]+\sum_{u'\in\mathrm{Nei}(u)\backslash i}\frac{P_{u'}}{k_{u'}}.\label{eq:walker_concentration_clustered_orig}
\end{equation}
We now replace every neighbor $u$ with an average node
of degree $\expv k$. In addition, we assume every node $i$ has
the same number of edges between neighbors, so $\kappa_{i}\equiv\kappa=C(\expv k-1)$.
This transforms Equation (\ref{eq:walker_concentration_clustered_orig})
into a self-consistent expression given by 
\[
P_{u}=\frac{1}{N\expv k}(\expv k-1-\kappa)+\kappa\frac{P_{u}}{\expv k},
\]
which is equivalent to
\[
P_{u}=\frac{1}{N}\left(1-\frac{1}{\expv k-C\big[\expv k-1\big]}\right).
\]
The decay rate is then
\[
\beta_{i}=\frac{k_{i}}{N\expv k}\left(1-\frac{1}{\expv k-C\big[\expv k-1\big]}\right).
\]
We therefore find that the global mean FPT of target node $i$ is
\[
\tau_{i}=\frac{N\expv k}{k_{i}}\left(\frac{1}{1-\Big[\expv k-C\big[\expv k-1\big]\Big]^{-1}}\right).
\]
In order to compute the pair--averaged FPT (i.e. the mean global mean FPT)
$\expv\tau=(1/N)\sum_{i}\tau_{i}$ for SSMH networks we need to compute the
mean inverse degree $\expv{k^{-1}}_{k>0}$, which can be numerically
intensive. We thus approximate this quantity by its second order Taylor expansion
around the mean $\expv k$, such that
\begin{align}
\expv{\frac{1}{k}}_{k>0}\approx\chi(\xi) & =\frac{1}{\expv k}+\frac{\expv{k^{2}}}{\expv{k}^{3}}+\Phi_{ER}\nonumber \\
 & =\frac{\expv{k^{2}}-\expv{k^{2}}_{ER}}{\expv{k}^{3}}+\expv{\frac{1}{k}}_{ER,k>0}.\label{eq:chi_xi}
\end{align}
Here, we also used the expressions for the transitivity $C$ in Equation (\ref{eq:clustering_coefficient}) and for the second moment of the degree distribution in Equation~(\ref{eq:second_moment}).
The constant $\Phi_{ER}$ contains the error made by considering networks
without nodes of $k=0$, which alters the degree distribution in comparison
to the second moment $\expv{k^{2}}$. We fix this issue by demanding
$\chi(\xi=B)=\expv{k^{-1}}_{ER}$, which yields
\[
\Phi_{ER}=\expv{\frac{1}{k}}_{ER,k>0}-\frac{1}{\expv k}-\frac{\expv{k^{2}}_{ER}}{\expv{k}^{3}}.
\]

\setcounter{equation}{0}
\renewcommand\theequation{E.\arabic{equation}}
\setcounter{figure}{0}
\renewcommand\thefigure{E.\arabic{figure}}

\section{Standard Watts--Strogatz small--world networks}
\label{app:mapping_to_WS_SW}

In this Appendix we describe how the SSMH and PLSW network models discussed in the main text can be mapped to a standard Watts--Strogatz small--world model, thus showing that the emergence of a minimal pair--averaged FPT at intermediate levels of a structural control parameter can be a generic feature in networks with some type of long--range links.

In the original small--world network model by Watts and Strogatz \cite{watts_collective_1998}, the structural control parameter that interpolates between a regular nearest--neighbor network and a randomized structure is the rewiring probability $p_r$. 
This is the probability with which one end of each link is switched from a neighboring node to any, randomly chosen, node in the network. 
For $p_r=0$, the network is a $k$--regular nearest--neighbor lattice, with $k$ an even integer. For $0<p_r<1$, each of the $k/2$ edges connecting every node $i$ to its $k/2$ right--side neighbors is rewired with probability $p_r$ to any other node $j$, chosen uniformly at random while avoiding self--connecting and duplicate edges. 
At $p_r=1$ all edges have been rewired. The resulting network is not structurally equivalent to a random graph model realization, however, with one of the differences being that every node has a degree greater or equal to $k/2$.

In order to obtain instead a variant of the Watts--Strogatz model that truly interpolates between a $k$--regular nearest--neighbor lattice and the random graph model, we define a connection probability distribution from which edges are drawn instead of rewired. 

As in Section \ref{sec:kleinberg_model}, we begin by defining the distance between two nodes with integer indices $i$ and $j$ as $n(i,j) = \mathrm{min}(|i-j|,N-|i-j|)$. Each pair of nodes within distance $n\leq k/2$ is then connected with a short--range probability $p_S$, while each pair of nodes with distance $n> k/2$ is connected with a long--range probability $p_L$. We require the mean degree to be constant, such that 
\begin{align}
	\label{eq:WS_norm_condition}
	k_S + k_L = p_S k + p_L (N-1-k) &= k
\end{align}
at all times. Since we are interested in having a controlled redistribution of the probability (and, consequently, of the edges) between short--range and long--range node pairs, we introduce a parameter $0\leq \beta \leq 1$ and set
\begin{align*}
	p_S &= p_0\\
	p_L &= \beta p_0.
\end{align*}
From here, using Equation~(\ref{eq:WS_norm_condition}) we find
\begin{align*}
	p_0 &= \frac {1}{1-\beta+\beta(N-1)/k}.
\end{align*}
Note that for $\beta=0$, we have $p_S = 1$ and $p_L = 0$, generating a $k$-regular nearest-neighbor lattice while for $\beta = 1$ we find $p_S = p_L = k/(N-1)$ which reduces the model to the Erdős--Rényi random graph model. In total, the connection probability is
\begin{align*}
	p(i,j) = \begin{cases}
    			 \frac {1}{1-\beta+\beta(N-1)/k}, &\mathrm{if}\ n(i,j)\leq k/2\\
                 \frac {\beta}{1-\beta+\beta(N-1)/k}, &\mathrm{otherwise}.
    		\end{cases}
\end{align*}
In order to make appropriate comparisons to the SSMH model and the PLSW model, we map the redistribution parameter $\beta$ to the structural control parameters $\xi$ and $\mu$. We do so by demanding that the expected short--range degree of the Watts--Strogatz model $k_S=kp_S$ is approximately equal to the short--range degree of the power--law small world model.
\begin{figure}
    \centering
    \includegraphics[width=3.375in]{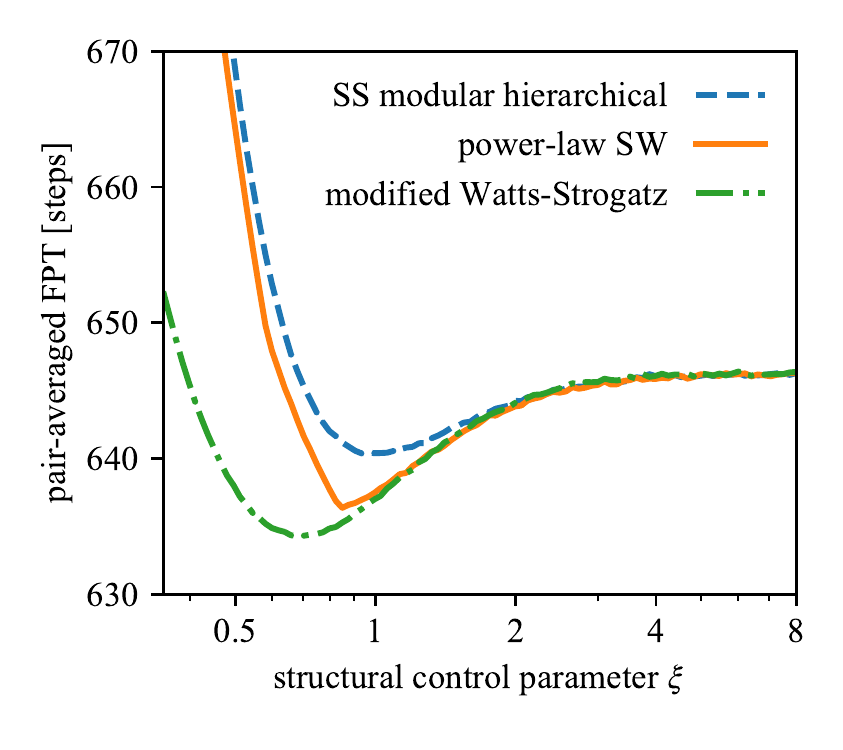}
    \caption{
        \label{fig:mhrn_kleinberg_smallworld}
        Pair--averaged first passage time for the SSMH, the PLSW and the modified Watts--Strogatz model with $B=8$, $L=3$, $\expv k = 10$, averaged over 10,000 realizations each. We observe that the presence of a minimum at intermediate levels of the structural control parameter $\xi$ is a common feature in all three models.
    }
\end{figure}
To this end, we approximate the power--law small world connection probability as being asymmetrical to the right and only approximately normed as 
\begin{align*}
	c(n) &\approx k\frac{n^{\mu-1}}{\int\limits_1^{N-1}dn\,n^{\mu-1}}\\
         &= \mu k\frac{n^{\mu-1}}{(N-1)^\mu-1}.
\end{align*}
In order for both models to produce a short--range mean degree of similar order, we then demand that
\begin{align*}
	kp_S = \int\limits_1^k dn\,c(n)
\end{align*}
and hence
\begin{align*}
	\tilde{\beta} = \frac{k}{N-1-k}\left[ \frac{(N-1)^\mu-1}{k^\mu - 1}-1\right].
\end{align*}
Since this result was obtained using approximations, we introduce a corrective factor and find
\begin{align*}
    \beta = \frac{k-1}{k}\tilde \beta = \frac{k-1}{N-1-k}\left[ \frac{(N-1)^\mu-1}{k^\mu - 1}-1\right]
\end{align*}
such that $\beta = 1$ for $\mu = 1$ and $\beta \rightarrow 0$ for $\mu\rightarrow\infty$. Note that in a similar manner the limit $\mu\rightarrow0$ yields
\begin{align*}
	\beta= \frac{k-1}{N-1-k}\frac{\log(N-1) - \log k}{\log k}.
\end{align*}

We compared our results for the pair--averaged FPT to both models in the main text. For parameters $B = 8$, $L=3$ ($N=8^3$, correspondingly), $\expv k = k = 10$ and a varying SSMH structural control parameter $\xi$, we computed the corresponding PLSW parameter $\mu$ and the Watts--Strogatz redistribution parameter $\beta$ and generated 10,000 single realizations of SSMH networks, PLSW and modified Watts--Strogatz networks. We subsequently computed the pair--averaged FPT as described in the main text and averaged over all realizations to obtain the curves shown in Figure~\ref{fig:mhrn_kleinberg_smallworld}. We observe that the minimum for the pair--averaged FPT emerges in all three cases. We have thus shown that this effect persists for all the tested network models where local clustering decreases while node connection heterogeneity increases.



\begin{thebibliography}{99}

\bibitem{fortunato_community_2010}
Fortunato, S. (2010)  Community detection in graphs. {\em Physics Reports},
  \textbf{486}(3{\textendash}5), 75--174.

\bibitem{lancichinetti_detecting_2009}
Lancichinetti, A., Fortunato, S. {\&} Kertesz, J. (2009)  Detecting the
  overlapping and hierarchical community structure of complex networks. {\em
  New Journal of Physics}, \textbf{11}(3), 033015.
arXiv: 0802.1218.

\bibitem{albert_statistical_2002}
Albert, R. {\&} Barabasi, A.-L. (2002)  Statistical mechanics of complex
  networks. {\em Reviews of Modern Physics}, \textbf{74}(1), 47--97.
arXiv: cond-mat/0106096.

\bibitem{newman_modularity_2006}
Newman, M. E.~J. (2006)  Modularity and community structure in networks. {\em
  PNAS}, \textbf{103}(23), 8577--8582.

\bibitem{sales-pardo_extracting_2007}
Sales-Pardo, M., Guimer{\`a}, R., Moreira, A.~A. {\&} Amaral, L. A.~N. (2007)
  Extracting the hierarchical organization of complex systems. {\em Proc Natl
  Acad Sci U S A}, \textbf{104}(39), 15224--15229.

\bibitem{clauset_hierarchical_2008}
Clauset, A., Moore, C. {\&} Newman, M. E.~J. (2008)  Hierarchical structure and
  the prediction of missing links in networks. {\em Nature},
  \textbf{453}(7191), 98--101.

\bibitem{peixoto_hierarchical_2014}
Peixoto, T.~P. (2014)  Hierarchical block structures and high-resolution model
  selection in large networks. {\em Phys. Rev. X}, \textbf{4}(1), 011047.

\bibitem{rosvall_multilevel_2011}
Rosvall, M. {\&} Bergstrom, C.~T. (2011)  Multilevel compression of random
  walks on networks reveals hierarchical organization in large integrated
  systems. {\em PLoS ONE}, \textbf{6}(4), e18209.

\bibitem{simon_architecture_1962}
Simon, H.~A. (1962)  The architecture of complexity. {\em Proc. Am. Philos.
  Soc.}, \textbf{106}(6), 467--482.

\bibitem{meunier_hierarchical_2009}
Meunier, D., Lambiotte, R., Fornito, A., Ersche, K.~D. {\&} Bullmore, E.~T.
  (2009)  Hierarchical modularity in human brain functional networks. {\em
  Front Neuroinformatics}, \textbf{3}.

\bibitem{meunier_modular_2010}
Meunier, D., Lambiotte, R. {\&} Bullmore, E.~T. (2010)  Modular and
  hierarchically modular organization of brain networks. {\em Front Neurosci},
  \textbf{4}, 200.

\bibitem{kaiser_limited_2010}
Kaiser, M. {\&} Simonotto, J. (2010)  Limited spreading: {How} hierarchical
  networks prevent the transition to the epileptic state. In Steyn-Ross, D.~A.
  {\&} Steyn-Ross, M., editors, {\em Modeling {Phase} {Transitions} in the
  {Brain}}, number~4 in Springer {Series} in {Computational} {Neuroscience},
  pages 99--116. Springer New York.

\bibitem{robinson_dynamical_2009}
Robinson, P.~A., Henderson, J.~A., Matar, E., Riley, P. {\&} Gray, R.~T. (2009)
   Dynamical reconnection and stability constraints on cortical network
  architecture. {\em Phys. Rev. Lett.}, \textbf{103}(10), 108104.

\bibitem{sarkar_spectral_2013}
Sarkar, S., Henderson, J.~A. {\&} Robinson, P.~A. (2013)  Spectral
  characterization of hierarchical network modularity and limits of modularity
  detection. {\em PLoS ONE}, \textbf{8}(1), e54383.

\bibitem{klimm_resolving_2014}
Klimm, F., Bassett, D.~S., Carlson, J.~M. {\&} Mucha, P.~J. (2014)  Resolving
  structural variability in network models and the brain. {\em PLOS
  Computational Biology}, \textbf{10}(3), e1003491.

\bibitem{ravasz_hierarchical_2002}
Ravasz, E., Somera, A.~L., Mongru, D.~A., Oltvai, Z.~N. {\&} Barab{\'a}si,
  A.-L. (2002)  Hierarchical organization of modularity in metabolic networks.
  {\em Science}, \textbf{297}(5586), 1551--1555.

\bibitem{barabasi_network_2004}
Barab{\'a}si, A.-L. {\&} Oltvai, Z.~N. (2004)  Network biology: understanding
  the cell's functional organization. {\em Nat Rev Genet}, \textbf{5}(2),
  101--113.

\bibitem{yerra_emergence_2005}
Yerra, B.~M. {\&} Levinson, D.~M. (2005)  The emergence of hierarchy in
  transportation networks. {\em Ann Reg Sci}, \textbf{39}(3), 541--553.

\bibitem{smith_hierarchical_2014}
Smith, C., Puzio, R.~S. {\&} Bergman, A. (2014)  Hierarchical network structure
  promotes dynamical robustness. {\em arXiv:1412.0709 [nlin, physics:physics,
  q-bio]}.
arXiv: 1412.0709.

\bibitem{webster_hierarchical_1979}
Webster, J.~R. (1979)  Hierarchical organization of ecosystems. In {\em
  Theoretical {Systems} {Ecology}}, pages 119--129. Academic Press.

\bibitem{arenas_synchronization_2006}
Arenas, A., D{\'i}az-Guilera, A. {\&} P{\'e}rez-Vicente, C.~J. (2006)
  Synchronization reveals topological scales in complex networks. {\em Phys.
  Rev. Lett.}, \textbf{96}(11), 114102.

\bibitem{pan_modular_2008}
Pan, R.~K. {\&} Sinha, S. (2008)  Modular networks with hierarchical
  organization: {The} dynamical implications of complex structure. {\em
  Pramana}, \textbf{71}(2), 331--340.
arXiv: 0903.1909.

\bibitem{rao_dynamic_2013}
Rao, V. S.~H. {\&} Durvasula, R. (2013) {\em Dynamic models of infectious
  diseases}, volume~2.
Springer Science \& Business Media.

\bibitem{watts_collective_1998}
Watts, D.~J. {\&} Strogatz, S.~H. (1998)  Collective dynamics of 'small-world'
  networks. {\em Nature}, \textbf{393}(6684), 440--442.

\bibitem{watts_identity_2002}
Watts, D.~J., Dodds, P.~S. {\&} Newman, M. E.~J. (2002)  Identity and search in
  social networks. {\em Science}, \textbf{296}(5571), 1302--1305.

\bibitem{travers_small_1967}
Travers, J. {\&} Milgram, S. (1967)  The small world problem. {\em Phychology
  Today}, \textbf{1}, 61--67.

\bibitem{kleinberg_small--world_2000}
Kleinberg, J. (2000)  The small--world phenomenon: {An} algorithmic
  perspective. In {\em Proceedings of the {Thirty}-second {Annual} {ACM}
  {Symposium} on {Theory} of {Computing}}, {STOC} '00, pages 163--170, New
  York, NY, USA. ACM.

\bibitem{newman_networks:_2010}
Newman, M. E.~J. (2010) {\em Networks: an introduction}.
Oxford University Press, Oxford ; New York.

\bibitem{bruggeman_berechnung_1935}
Bruggeman, D. A.~G. (1935)  Berechnung verschiedener physikalischer
  {Konstanten} von heterogenen {Substanzen}. {I}.
  {Dielektrizit{\"a}tskonstanten} und {Leitf{\"a}higkeiten} der
  {Mischk{\"o}rper} aus isotropen {Substanzen}. {\em Annalen der Physik},
  \textbf{416}(7), 636--664.

\bibitem{thiel_effective-medium_2016}
Thiel, F. {\&} Sokolov, I.~M. (2016)  Effective-medium approximation for
  lattice random walks with long-range jumps. {\em Phys. Rev. E},
  \textbf{94}(1), 012135.

\bibitem{batagelj_efficient_2005}
Batagelj, V. {\&} Brandes, U. (2005)  Efficient generation of large random
  networks. {\em Phys. Rev. E}, \textbf{71}(3), 036113.

\bibitem{maier_cmhrn_2018}
Maier, B.~F. (2018)  {cMHRN} - {A} {C}++/{Python}/{MATLAB} package to generate
  {SSMH} and {PLSW} networks in a fast manner,
  https://github.com/benmaier/{cMHRN}. .

\bibitem{lin_mean_2012}
Lin, Y., Julaiti, A. {\&} Zhang, Z. (2012)  Mean first-passage time for random
  walks in general graphs with a deep trap. {\em J Chem Phys},
  \textbf{137}(12), 124104.

\bibitem{maier_cnetworkdiff_2017}
Maier, B.~F. (2017)  {cNetworkDiff} - {A} {C}++/{Python}/{MATLAB} package for
  random walk simulations on networks,
  https://github.com/benmaier/{cNetworkDiff}. .

\bibitem{sood_first-passage_2005}
Sood, V., Redner, S. {\&} ben-Avraham, D. (2005)  First-passage properties of
  the {Erd{\H o}s}--{Renyi} random graph. {\em J. Phys. A: Math. Gen.},
  \textbf{38}(1), 109.

\bibitem{maier_cover_2017}
Maier, B.~F. {\&} Brockmann, D. (2017)  Cover time for random walks on
  arbitrary complex networks. {\em Phys. Rev. E}, \textbf{96}(4), 042307.

\bibitem{lau_asymptotic_2010}
Lau, H.~W. {\&} Szeto, K.~Y. (2010)  Asymptotic analysis of first passage time
  in complex networks. {\em EPL}, \textbf{90}(4), 40005.

\bibitem{barrat_dynamical_2008}
Barrat, A., Barth{\'e}lemy, M. {\&} Vespignani, A. (2008) {\em Dynamical
  {Processes} on {Complex} {Networks}}.
Cambridge University Press, 1 edition.

\end{thebibliography}
%








\end{document}